\documentclass[aps,twocolumn,showpacs,preprintnumbers,nofootinbib,prd,10pt,superscriptaddress,longbibliography,firstinits]{revtex4-1}

\makeatletter
\def\l@subsubsection#1#2{}
\def\l@subsubsubsection#1#2{}
\makeatother

\usepackage{graphicx,amssymb,amsmath,amsthm,amsfonts,fixmath}
\usepackage{comment}
\usepackage{mathrsfs}
\usepackage{mathtools}
\usepackage[usenames]{color}
\usepackage{epstopdf}
\usepackage{bm}
\usepackage{dcolumn}
\usepackage{float}
\usepackage{rotating}
\usepackage{longtable}

\usepackage{enumerate}
\usepackage{tensor,multirow}
\usepackage{url}
\usepackage[dvipsnames]{xcolor}
\usepackage[unicode]{hyperref}
\hypersetup{colorlinks=true, citecolor=MidnightBlue,
            linkcolor=Maroon, urlcolor=MidnightBlue, linktocpage=true}
\usepackage{orcidlink}

\begin{document}

\title{Excitation factors and exceptional points of Kerr-de Sitter quasinormal modes}

\author{Dario Rossi
\orcidlink{0009-0002-9358-342X}}
\email{dario.rossi@phd.unipi.it}
\affiliation{Dipartimento di Fisica, Università di Pisa, Largo B. Pontecorvo 3, 56127 Pisa, Italy}
\affiliation{INFN, Sezione di Pisa, Largo B. Pontecorvo 3, 56127 Pisa, Italy}

\author{Naritaka Oshita
\orcidlink{0000-0002-8799-1382}}
\email{oshita@phys.kindai.ac.jp}
\affiliation{Department of Physics, Kindai University, Osaka 577-8502, Japan}
\affiliation{RIKEN iTHEMS, Wako, Saitama, 351-0198, Japan}

\author{Leonardo Gualtieri}
\email{leonardo.gualtieri@unipi.it}
\affiliation{Dipartimento di Fisica, Università di Pisa, Largo B. Pontecorvo 3, 56127 Pisa, Italy}
\affiliation{INFN, Sezione di Pisa, Largo B. Pontecorvo 3, 56127 Pisa, Italy}
\email{leonardo.gualtieri@unipi.it}

\author{Emanuele Berti}
\affiliation{Department of Physics and Astronomy, Johns Hopkins University,
3400 N. Charles Street, Baltimore, Maryland, 21218, USA}
\email{berti@jhu.edu}

\newcommand{\sqAlpha}{\sqrt{\alpha}}
\newcommand{\rmin}{r_-}
\newcommand{\rmm}{r_-'}
\newcommand{\rp}{r_+}
\newcommand{\rpp}{r_+'}
\newcommand{\dr}[1]{\textcolor{blue}{\sf[DR: #1]}}
\newcommand{\leo}[1]{\textcolor{brown}{\sf[LG: #1]}}
\newcommand{\no}[1]{\textcolor{orange}{\sf[NO: #1]}}
\newcommand{\eb}[1]{\textcolor{red}{\sf[EB: #1]}}

\begin{abstract}
  We compute the excitation factors of Kerr-de Sitter black hole quasinormal modes using two different techniques: a Heun function representation and an expansion of the solution in terms of hypergeometric functions. Exceptional points at which the complex frequencies of different overtones coincide are a generic feature of the Kerr-de Sitter quasinormal mode spectrum. Near exceptional points, the modes exhibit a ``hysteresis phenomenon'' that was previously found for Kerr black holes perturbed by massive fields and the absolute value of the excitation factors of the individual modes is significantly enhanced, while they acquire nearly opposite phases. We extrapolate the excitation factors to small values of the cosmological constant and find good agreement with previous calculations in the Kerr limit.
\end{abstract}

\maketitle

\section{Introduction}\label{sec:intro}
The detection of gravitational waves (GWs) from the merger of two black holes (BHs)~\cite{LIGOScientific:2016aoc} has opened a new avenue for testing gravity in the strong-field regime~\cite{LIGOScientific:2026qni, LIGOScientific:2026fcf, LIGOScientific:2026wpt}.
In this context, BH spectroscopy -- the idea of measuring multiple quasinormal mode (QNM) frequencies and testing their consistency with the predictions of general relativity -- is a promising approach to perform precision tests of gravity and of the nature of compact objects (see e.g.~\cite{Kokkotas:1999bd, Berti:2009kk, Berti:2025hly} for reviews).

The complex structure of the Kerr BH spectrum has been studied extensively because of its observational interest for BH spectroscopy.
In comparison, there is relatively little work on the QNM spectrum of the Kerr-de Sitter (KdS) background, that can be written in Boyer--Lindquist coordinates as
\begin{equation}
\begin{split}
    ds^2=&-\frac{{\rm \Delta}_r(r)}{(1+\alpha)^2 \rho^2}\left(dt -a \sin^2 \theta d\phi \right)^2 \\& +\frac{{\rm \Delta}_\theta \sin^2\theta}{(1+\alpha)^2 \rho^2}\left\{a dt - (r^2 + a^2) d\phi\right\}^2\\
    &+\frac{\rho^2}{{\rm \Delta}_r(r)}dr^2+\frac{\rho^2}{{\rm \Delta}_\theta}d\theta^2\,,
    \label{eq:KdS}
\end{split}
\end{equation}
where
\begin{align}
    &\rho^2 \equiv r^2+a^2\cos^2{\theta}\,, \nonumber\\
    &{\rm \Delta}_r(r)\equiv(r^2+a^2)\left(1-\frac{{\rm \Lambda}}{3}r^2\right)-2 M r \nonumber\\
    & \qquad =-\frac{\rm \Lambda}{3}(r-\rmm)(r-\rmin)(r-\rp)(r-\rpp)\,, \nonumber\\
    &{\rm \Delta}_\theta\equiv1 + \alpha \cos^2\theta\,. \nonumber
\end{align}
Here $M$ is the BH mass, $a$ is the angular momentum per unit mass, $\rm \Lambda$ is the cosmological constant, and $\alpha \equiv \frac{{\rm \Lambda}}{3}a^2 \geq0$. 
The radii of the event horizon, $\rp$, and of the cosmological horizon, $\rpp$, are determined by the four roots of ${\rm \Delta}_r(r)=0$, labeled so that $\rmm<\rmin<\rp<\rpp$; $\rmin$ corresponds to the inner horizon, while $\rmm$ is always negative, and thus nonphysical. It is also useful to define a tortoise coordinate $r_\star$ such that $dr_\star/dr=(1+\alpha)(r^2+a^2)/{\rm \Delta}_{r}$~\cite{Yoshida:2010zzb,Motohashi:2021zyv}. Note that $r_\star\to-\infty$ at the event horizon, and $r_\star\to+\infty$ at the cosmological horizon.

The perturbations of the KdS background can be studied using the Newman-Penrose framework~\cite{Newman:1966ub}, and are described by a generalization of the Teukolsky equation~\cite{Teukolsky:1973ha}. The Weyl curvature scalar $\psi_4$ can be separated in angular and radial components as
\begin{equation}
    \label{eq:newman_penrose}
    \begin{aligned}
    \psi_4&=(r-ia \cos\theta)^{-4}\\ 
    & \times\int d\omega e^{-i \omega_{\ell mn}t}\sum_{\ell m}e^{im\varphi}\frac{S_{\ell m}(\omega,\theta)}{\sqrt{2\pi}}R_{\ell m}(\omega,r)\,,
    \end{aligned}
\end{equation}
where $\omega_{\ell mn}$ are the QNM frequencies, and $(\ell,\,m,\,n)$ denote the angular, azimuthal, and overtone numbers, respectively. The spin-weighted spheroidal harmonics $S_{\ell m}$ and the Teukolsky radial variable $R_{\ell m}$ satisfy generalizations of the Teukolsky perturbation equations in the KdS background that were derived in Refs.~\cite{Khanal:1983vb,Chambers:1994ap}.

Suzuki, Takasugi and Umetsu~(STU)~\cite{Suzuki:1998vy,Suzuki:1999nn,Suzuki:1999pa} generalized the work by Mano, Suzuki and Takasugi~(MST)~\cite{Mano:1996vt,Mano:1996mf,Mano:1996gn} to show that both the angular and radial equations are special cases of the Heun equation (see also~\cite{Batic:2007it,Hatsuda:2020sbn,Motohashi:2021zyv}). The KdS QNM spectrum was studied by Moss and Norman~\cite{Moss:2001ga}, and more extensively by Yoshida, Uchikata and Futamase~\cite{Yoshida:2010zzb} (see also~\cite{Dyatlov:2010hq,Dyatlov:2011jd} for more rigorous mathematical work, \cite{Tattersall:2018axd} for results in the slow-rotation limit, \cite{Novaes:2018fry} for a study using the accessory parameter expansion of Heun's equation, and~\cite{Dias:2018ynt,Casals:2020uxa,Hintz:2021rbv,Casals:2021ugr,Davey:2024xvd} for work on KdS QNMs in the context of strong cosmic censorship).

For observational purposes, a crucial property of QNMs is their amplitude, quantified by the excitation factors (EFs) $E_{\ell m n}$~\cite{Leaver:1986gd,Sun:1988tz,Andersson:1995zk,Glampedakis:2001js,Glampedakis:2003dn,Berti:2006wq,Zhang:2013ksa,Oshita:2021iyn,Motohashi:2024fwt,Oshita:2025ibu,Kubota:2025hjk}.
The GW strain amplitude $h(t,r)=h_+(t,r)+ih_\times(t,r)$
can be written in terms of the EFs as
\begin{equation}
\label{eq:KdSEF}
  h(t,r)=-\frac{2}{r}\frac{e^{i m \varphi}}{\sqrt{2\pi}}\sum_{\ell m n} E_{\ell m n} T_{\ell m n}S_{\ell m n}(\theta) e ^{-i\omega_{\ell m n}(t-r_\star)}\,,
\end{equation}
where $S_{\ell m n}(\theta)=S_{\ell m}(\omega_{\ell m n},\theta)$, $T_{\ell mn}$ are properly defined integrals of the source exciting the perturbations, and $E_{\ell m n}$ denotes the EFs of the strain in terms of the outgoing and ingoing wave amplitudes $A_{\ell m }^{(\text{ref})}$ and $A^{(\rm inc)}_{\ell m}$~\cite{Oshita:2021iyn,Oshita:2025ibu} (see Eq.~\eqref{eq:excitation_factor} below).
Another common definition of the QNM excitation is in terms of the quantities $B_{\ell m n} \equiv E_{\ell m n} \omega^2_{\ell mn}$, which are associated with the amplitude of $\psi_4$~\cite{Berti:2006wq,Motohashi:2024fwt,Kubota:2025hjk}.

The EFs of the KdS spacetime, $E_{\ell m n}$, have recently been computed using the Heun formulation of the perturbation equations~\cite{Oshita:2021iyn,Oshita:2025ibu}. 
Here we present a more comprehensive study of KdS QNMs and EFs, with two main motivations.

The first is recent progress in the understanding of the structure of the Green's function.
The importance of the excitation factors in asymptotically flat spacetimes has been known for a long time (see e.g.~\cite{Leaver:1986gd,Sun:1988tz,Andersson:1995zk,Andersson:1996cm,Glampedakis:2001js,Glampedakis:2003dn,Berti:2006wq,Zhang:2013ksa,Oshita:2022pkc,Oshita:2021iyn,Oshita:2024wgt,DeAmicis:2025xuh,Kuntz:2025gdq,Oshita:2025ibu,Kubota:2025hjk,Oshita:2026vxh,Kubota:2026hdv,DellaRocca:2025zbe,Kuntz:2025gdq,DeAmicis:2026tus,Aruquipa:2026tga,Rosato:2026moe,DeAmicis:2026wqd,Kuntz:2026xep,Su:2026fvj}).
Recent work showed that Green’s functions of asymptotically de Sitter BH spacetimes can be expressed as a convergent mode sum everywhere in spacetime: at late times this sum involves QNMs, while at early times it involves Matsubara (or Euclidean) modes, with the two regions being separated by light cone scattering from the BH potential~\cite{Arnaudo:2025uos,Arnaudo:2025kit,Arnaudo:2026tcy}. The systematic evaluation of EFs for KdS QNMs is a necessary first step to extend this analysis to the rotating case (see e.g.~\cite{Motohashi:2026mbn} for a study of the contribution of QNMs and Matsubara modes in the Kerr case).

The second motivation is related to avoided crossings (ACs), that have attracted significant interest~\cite{Dias:2021yju,Davey:2022vyx,Dias:2022oqm,Davey:2023fin,Motohashi:2024fwt,Cavalcante:2024kmy,Cavalcante:2025abr,Oshita:2025ibu,Lo:2025njp,Yang:2025dbn,Kubota:2025hjk,Cao:2025afs,PanossoMacedo:2025xnf,Xie:2025qoi,Nakamoto:2026lyo,Cheng:2026gxu,Zhou:2026xzf,Kubota:2026hdv,Imafuku:2026rpn,Cavalcante:2026vgr} because they can lead to a large amplification of the relevant QNM amplitudes~\cite{Motohashi:2024fwt} with destructive interference~\cite{Oshita:2025ibu} near ACs.
They can exhibit beating-like phenomena~\cite{Yang:2025dbn}, and this may have interesting observational signatures (see e.g.~\cite{Takahashi:2025uwo,Imafuku:2026rpn}).
For BH perturbations characterized by at least two parameters, ACs can turn into actual mode crossings, i.e., {\it exceptional points} (EPs). Among the simplest additional parameters one can consider the mass of the perturbing field~\cite{Cavalcante:2024swt,Cavalcante:2024kmy,Cavalcante:2025abr}, the electric charge in Kerr-Newman BHs~\cite{Cavalcante:2026vgr}, or (as we do in this work) the cosmological constant in KdS BHs~\cite{Oshita:2021iyn,Oshita:2025ibu}.

In Ref.~\cite{Oshita:2025ibu}, one EP was identified in the QNM spectrum of KdS BHs. This EP corresponds to a crossing between the $n=5$ and $n=6$ overtones of perturbations with $(\ell,\,m)=(2,\,2)$, that occurs for $M^2{\rm \Lambda}\simeq0.0085$ and $a/M\simeq0.896$. The corresponding EFs~\cite{Leaver:1986gd,Sun:1988tz,Andersson:1995zk,Glampedakis:2001js,Glampedakis:2003dn,Berti:2006wq,Zhang:2013ksa,Oshita:2021iyn} are enhanced and (in principle) diverge at the EP, but the amplitude of the two superposed QNMs remains finite, as the amplification occurs in a destructive manner~\cite{Oshita:2025ibu,Yang:2025dbn}.

In this paper, we perform a systematic study of the QNM frequencies and EFs of KdS BHs, finding that EPs are a generic property of the spectrum. In addition, we compute the EFs near the EPs. 

The numerical calculation of the EFs of KdS QNMs is a delicate task. For this reason, we compute the EFs using two independent methods. The first (also employed in~\cite{Oshita:2021iyn,Oshita:2025ibu}) is based on the Heun representation of the solution of the generalized Teukolsky equation using the built-in \texttt{HeunG} functions of \textsc{Mathematica}~\cite{Hatsuda:2020sbn}; the second, derived in this work, is a generalization of the MST/STU method~\cite{Mano:1996vt,Mano:1996mf,Mano:1996gn,Suzuki:1998vy,Suzuki:1999nn,Suzuki:1999pa}. We validate the two methods by showing that they give comparable results within the numerical errors. 

Finally, we analyze the behavior of the QNMs in the vicinity of the EPs in the $(a/M,\,M^2 {\rm \Lambda})$ parameter space. We find a hysteresis phenomenon analogous to that found for rotating BHs perturbed by a massive scalar field~\cite{Cavalcante:2024kmy,Cavalcante:2025abr}, or for Kerr-Newman BHs perturbed by a scalar field~\cite{Cavalcante:2026vgr}: along a closed trajectory in the $(a/M,\,M^2 {\rm \Lambda})$ Riemann plane around the EP, the two QNMs ``trade places'' due to a branch cut.

The plan of the paper is as follows.
In Sec.~\ref{sec:theory} we present the perturbation equations in the KdS background and define the QNM EFs. 
In Sec.~\ref{sec:KdS_QNM_spectrum} we investigate the QNM spectrum of KdS BHs and the associated EPs, and we discuss the phenomenon of hysteresis in the vicinity of EPs.
In Sec.~\ref{sec:excitation_factors} we compute and discuss the EFs of KdS QNMs, and their behavior near EPs. Finally, in Sec.~\ref{sec:concl} we draw our conclusions. Throughout the paper we use geometrical units ($G=c=1$).

\section{Perturbations equations and excitation factors}
\label{sec:theory}

The generalized Teukolsky equations for spin-$s$ perturbations of the KdS spacetime in vacuum can be separated into an angular equation~\cite{Suzuki:1998vy}
\begin{equation}
\label{eq:angular_Teukolsky}
    \begin{aligned}
       &\Bigg\{ \frac{\partial }{\partial x}\left(1+\alpha  x^2\right) \left(1-x^2\right) \frac{\partial}{\partial x}+\lambda -s (1-\alpha )-2 \alpha  x^2+\\ &\frac{(1+\alpha )^2}{1+\alpha  x^2}\bigg(a^2 \omega ^2 x^2-2 a \omega  s x-a^2 \omega ^2+\\ & 2 a \omega  m+\frac{(4 \alpha ) s m x}{1+\alpha }-\frac{(m+s x)^2}{1-x^2}\bigg)\Bigg\} S_{\ell m}(x) = 0\,,
    \end{aligned}
\end{equation}
with $x\equiv \cos{\theta}$, and a radial equation
\begin{equation}
\label{eq:radial_Teukolsky}
    \begin{aligned}
        &{\rm \Delta}_r^{-s}(r) \frac{\partial}{\partial r}\left({\rm \Delta}_r^{s+1}(r) \frac{d}{dr}R_{\ell m}(r)\right)+\Big\{\frac{1}{{\rm \Delta}_r(r)}\Big((1+\alpha )^2 K(r)^2-\\ &-i s (1+\alpha ) K(r) \frac{d {\rm \Delta}_r(r)}{d r}\Big)+\Big(4 i s (1+\alpha ) \omega  r-\\ &-\frac{2 \alpha}{a^2} (s+1) (2 s+1) r^2+2 s (1-\alpha )-\lambda \Big)\Big\} R_{\ell m}(r)=0\,,
    \end{aligned}
\end{equation}
where $K(r) \equiv \omega (r^2 + a^2) - a m$, and $\lambda_{\ell m n}=\lambda$ (not to be confused with the cosmological constant $\rm \Lambda$) denotes the angular separation constant.

\subsection{Solutions of the perturbation equations}
By definition, the QNMs are solutions of the perturbation equations~\eqref{eq:angular_Teukolsky} and~\eqref{eq:radial_Teukolsky} satisfying ingoing boundary conditions at the BH event horizon and outgoing boundary conditions at the cosmological horizon.
The general solution of the radial Teukolsky equation~\eqref{eq:radial_Teukolsky} can be expressed as a linear combination of two independent solutions satisfying ingoing boundary conditions at the BH event horizon, $R^{(\text{in})}_{\ell m}(r)$, and outgoing boundary conditions at the cosmological horizon, $R^{(\text{up})}_{\ell m}(r)$~\cite{Motohashi:2021zyv}. Their asymptotic behaviors are
\begin{align}
\label{eq:asymptotic_solutions_1}
    &R^{(\text{in})}_{\ell m}(r)\simeq
    \begin{cases}
        A_{\ell m}^{(\rm trans)}{\rm\Delta}_r^{-B_1-s}\,, &(r\to \rp)\,,\\
        A_{\ell m}^{(\rm ref)}{\rm \Delta}_r^{B_2}+A_{\ell m}^{(\rm inc)}{\rm \Delta}_r^{-B_2-s}\,, & (r\to \rpp)\,,
    \end{cases}\\
    &R^{(\text{up})}_{\ell m}(r)\simeq
    \begin{cases}
        C_{\ell m}^{(\rm up)}{\rm \Delta}_r^{B_1}+C_{\ell m}^{(\rm ref)}{\rm \Delta}_r^{-B_1-s}\,, &(r\to \rp)\,,\\
        C_{\ell m}^{(\rm trans)}{\rm \Delta}_r^{B_2}\,, & (r\to \rpp)\,,
    \end{cases}
\label{eq:asymptotic_solutions_2}
\end{align}
with
\begin{subequations}
\label{eq:B1_B2}
\begin{align}
    B_1&= B(\rp)\,, \\
    B_2&= B(\rpp)\,,\label{eq:defB2} \\
    B(r)&=i  \frac{(\alpha +1) K(r)}{{\rm \Delta}_r'(r)}\,.
\end{align}
\end{subequations}
Note that~\cite{Hatsuda:2020sbn}
\begin{align}
    \begin{cases}
    \label{eq:EH_asymptotic_R1}
    {\rm\Delta}_r^{B_1}
        \simeq (r-\rp)^{i\omega \rp^{2}/{\rm \Delta}'_r(\rp)},\\[2mm]
    {\rm\Delta}_r^{-B_1-s}
        \simeq (r-\rp)^{-s-i\omega \rp^2/{\rm \Delta}'_r(\rp)},
    \end{cases}
    &\quad (r\to \rp)\,,
    \\[3mm]
    \begin{cases}
    {\rm \Delta}_r^{B_2}
        \simeq (r-\rpp)^{i\omega \rpp^{\,2}/{\rm \Delta}'_r(\rpp)},\\[2mm]
    {\rm \Delta}_r^{-B_2-s}
        \simeq (r-\rpp)^{-s-i\omega \rpp^{\,2}/{\rm \Delta}'_r(\rpp)},
    \end{cases}
    &\quad (r\to \rpp)\, .
    \label{eq:EH_asymptotic_R1_zto1}
\end{align}
In Eq.~\eqref{eq:asymptotic_solutions_1}, the amplitudes $A^{(\text{ref})}_{\ell m}$ and $A^{(\text{inc})}_{\ell m}$ are the coefficients of the outgoing and ingoing modes at the cosmological horizon in $R^{(\text{in)}}_{\ell m}(r)$; the amplitudes $C^{(\text{up})}_{\ell m}$ and $C^{(\text{ref})}_{\ell m}$ in Eq.~\eqref{eq:asymptotic_solutions_2} are the coefficients of the outgoing and ingoing modes at the BH outer horizon in $R^{(\text{up})}_{\,\ell m}(r)$. 
All of these amplitudes depend on the frequency $\omega$. One can also write the solutions
\begin{align}
\label{eq:asymptotic_solutions_3}
    &R^{(\text{out})}_{\ell m}(r)={\rm \Delta}_r^{s}(R^{(\text{in})}_{\ell m}(r))^*|_{s\to-s}\,, \\&R^{(\text{down})}_{\ell m}(r)={\rm \Delta}_r^{s}(R^{(\text{up})}_{\ell m}(r))^*|_{s\to-s}\,,
\label{eq:asymptotic_solutions_4}
\end{align}
one of which ($R^{(\text{out})}_{\ell m}(r)$) satisfies outgoing boundary conditions at the BH event horizon, while the other ($R^{(\text{down})}_{\ell m}(r)$) satisfies ingoing boundary conditions at the cosmological horizon. Two of the four solutions~\eqref{eq:asymptotic_solutions_1}, \eqref{eq:asymptotic_solutions_2}, \eqref{eq:asymptotic_solutions_3}, \eqref{eq:asymptotic_solutions_4} are a set of independent solutions for Eq.~\eqref{eq:radial_Teukolsky}.

As follows from the definition above, the QNM frequencies $\omega_{\ell m n}$ can be found by imposing $A_{\ell m}^{(\text{inc})}(\omega)=0$ from Eq.~\eqref{eq:asymptotic_solutions_1}. 
We compute the QNM frequencies $\omega_{\ell m n}$ with Leaver's continued fraction method~\cite{Leaver:1985ax,Suzuki:1998vy,Yoshida:2010zzb} (see Appendix~\ref{appendix:continued_fraction_method} for details). To validate our results, we also employ a second method based on the representation of the solutions of Eq.~\eqref{eq:radial_Teukolsky} through the built-in Heun functions of \textsc{Mathematica}~\cite{Hatsuda:2020sbn,Oshita:2025ibu}. 

In order to compute the wave amplitudes $A_{\ell m}$ and QNM frequencies $\omega_{\ell m n}$ using these two methods, it is useful to rewrite the radial Teukolsky equation~\eqref{eq:radial_Teukolsky} as a generic Heun differential equation~\cite{HeunDiffEq} of the form
\begin{equation}
\label{eq:HeunEq}
\begin{split}
    &\Bigg\{\frac{\partial ^2}{\partial z^2} \, + \,  \left(\frac{\gamma }{z}+\frac{\delta }{z-1}+\frac{\epsilon }{z-z_a}\right)\frac{\partial}{\partial z}\, + \, \\
    & \quad +\frac{\alpha \beta z-q}{z (z-1) (z-z_a)}\Bigg\}y_{\ell m}(z) =0\,,
\end{split}
\end{equation}
where the expressions for the parameters $\gamma$, $\delta$, $\epsilon$, $\alpha$, $\beta$, $q$ are provided in Appendix~\ref{appendix:continued_fraction_method} to improve readability.

The Heun equation can be obtained by performing the following transformation of the radial function:
\begin{equation}
\label{eq:radial_eigenfunction}
\begin{split}
    R_{\ell m}(z)=&z^{B_1}(z-1)^{B_2}\left(z-z_a\right)^{B_3}\left(z-z_{\infty}\right)^{2s+1} y_{\ell m}(z)\,,
\end{split}
\end{equation}
where
\begin{equation}
\label{eq:z_variable}
    z(r)=\frac{(\rpp-\rmin)(r-\rp)}{(\rpp-\rp)(r-\rmin)}\,
\end{equation}
is a new radial coordinate that maps the BH event horizon into $z=0$ and the cosmological horizon into $z=1$, $z_a=z(\rmm)$, $z_\infty=\frac{(\rpp-\rmin)}{(\rpp-\rp)}$, $B_3=B(\rmm)$, and $B_1$ and $B_2$ were defined in Eq.~\eqref{eq:B1_B2}.
There are two independent solutions around each regular singular point of Eq.~\eqref{eq:HeunEq}~\cite{HeunDiffEq}.

We will express the solution that satisfies ingoing boundary conditions at the BH event horizon as a linear combination of the two independent solutions satisfying ingoing and outgoing boundary conditions at the cosmological horizon. Since
$z\propto{\rm \Delta}_r(r)$ for $z\to0$~\cite{Motohashi:2021zyv}, we select the solution of Eq.~\eqref{eq:HeunEq} that goes like $\sim z^{-2B_1-s}$ around $z=0$, so that the function $R^{(\text{in})}_{\ell m}$ has the requested behavior near the BH event horizon: $R^{(\text{in})}_{\ell m}\sim{\rm\Delta}_r^{-B_1-s}$, as in Eq.~\eqref{eq:EH_asymptotic_R1}.
Denote by $y^{(\text{up})}_{\ell m}$ and $y^{(\text{down})}_{\ell m}$ the two independent solutions around the singular point $z=1$ of Eq.~\eqref{eq:HeunEq}. In the limit $z\to1$ their behavior is $y^{(\text{up})}_{\ell m}\simeq(z-1)^0$ and $y^{(\text{down})}_{\ell m}\simeq(z-1)^{-2B_2-s}$~\cite{Hatsuda:2020sbn}. Using the fact that $1-z\propto{\rm\Delta}_r(r)$ in the limit $r\to\rpp$~\cite{Motohashi:2021zyv}, from Eq.~\eqref{eq:EH_asymptotic_R1_zto1} and Eq.~\eqref{eq:RinBH} we see that 
$y^{(\text{up})}_{\ell m}$ ($y^{(\text{down})}_{\ell m}$) is associated to the asymptotically outgoing (ingoing) solution at the cosmological horizon.
The homogeneous solution of the generalized Teukolsky equation~\eqref{eq:radial_Teukolsky} satisfying ingoing boundary conditions at the event horizon, $R^{(\text{in})}_{\ell m}$ of Eq.~\eqref{eq:asymptotic_solutions_1}, can thus be written as
\begin{equation}
\begin{aligned}
    R^{(\text{in})}_{\ell m}(z)&=z^{B_1}(z-1)^{B_2}(z-z_a)^{B_3}(z-z_{\infty})^{2s+1}\\ & \times \left[c^{(\text{out})}_{\ell m}(\omega)\,y^{(\text{up})}_{\ell m}(z)+c^{(\text{in})}_{\ell m}(\omega)\,y^{(\text{down})}_{\ell m}(z)\right]\, ,
\label{eq:RinBH}
\end{aligned}
\end{equation}
where $c^{(\text{out})}_{\ell m}(\omega)\propto A^{(\text{ref})}_{\ell m}$ and $c^{(\text{in})}_{\ell m}(\omega)\propto A^{(\text{inc})}_{\ell m}$.
In practice, when we use \textsc{Mathematica}'s implementation of the Heun functions to compute the QNM frequencies, the quantity $c^{(\text{in})}_{\ell m}(\omega)$ is computed in terms of the Wronskian of the Heun equation, and the QNM frequencies are found by imposing $c^{(\text{in})}_{\ell m}(\omega_{\ell m n})=0$~\cite{Hatsuda:2020sbn,Oshita:2021iyn,Motohashi:2021zyv}.

\subsection{Excitation factors}
\label{subsec:efKdS}
When a BH is linearly perturbed by a material source, multiple QNMs are excited with different amplitudes.
Each QNM amplitude involves an integral $T_{\ell m n}$ over the source of the perturbation as well as the intrinsic excitability of each QNM, which is quantified by the EFs $E_{\ell m n}$~\cite{Berti:2006wq} defined in Eq.~\eqref{eq:excitation_factor}.

Ideally, we would like to normalize the ingoing and outgoing amplitudes (and thus the EFs) in a way that is consistent with the standard normalization of ingoing and outgoing amplitudes in the Kerr spacetime~\cite{Teukolsky:1973ha,Berti:2006wq}. The implementation of this choice is nontrivial, since the asymptotic behavior of the Teukolsky equation near infinity in Kerr spacetime is {\it different} from the behavior of $R^{\text{(in)}}$, $R^{\text{(up)}}$ near the cosmological horizon. However, as noted in~\cite{Oshita:2021iyn}, when $M^2 {\rm \Lambda}\ll1$, there is a region $r_+\ll r\ll r'_+$ in which the ingoing and outgoing solutions of the generalized Teukolsky equation in the KdS background behave like the solutions of the Teukolsky equation in the Kerr background. In this region, it is possible to extract ingoing and outgoing amplitudes in the KdS spacetime with a normalization consistent with that of the Kerr amplitudes. With this normalization, as we discuss below, the KdS EFs reduce to those of Kerr spacetime for $M^2 {\rm \Lambda}\to0$ (see also Appendix~\ref{appendix:EF_Kerr_limit}). This choice leads to the following normalization of the ingoing and outgoing amplitudes~\cite{Oshita:2021iyn}:
\begin{align}
    A^{(\rm inc)}_{\ell m}&=c_{\ell m}^{(\text{in})}\mathcal{Q}_{\rm in}\,2^{2s} \frac{a}{\sqrt{\alpha}}\,,\nonumber\\
    A^{(\rm ref)}_{\ell m}&=c_{\ell m}^{(\text{out})}\mathcal{Q}_{\rm out}\, \left(\frac{\alpha}{a^2}\right)^{-2s-1}\,,
    \label{eq:amplnorm}
\end{align}
where

\begin{align}
\begin{split}
    \mathcal{Q}_{\rm out} &=
    (-1)^{B_2}(1-z_a)^{B_3}(1-z_{\infty})^{2s+1} \\  
    & \quad \times \left(\frac{(r_+-r_{-})(r'_+-r'_{-})}{(r'_{+}-r_-)(r'_+-\rp)}\right)^{B_2}\,, 
    \end{split}\\
    \begin{split}
    \mathcal{Q}_{\rm in} &=
    (-1)^{B_2}(1-z_a)^{B_3}(1-z_{\infty})^{2s+1} \\
    & \quad \times \left(\frac{(r_+-r_{-})(r'_+-r'_{-})}{(r'_{+}-r_-)(r'_+-\rp)}\right)^{-s-B_2}\,,
    \end{split}
\end{align} 
which are obtained from Eq.~\eqref{eq:RinBH} in the $z\to1$ regime.

The EFs can be defined in terms of the amplitudes~\eqref{eq:amplnorm}:
\begin{equation}
\label{eq:excitation_factor}
    E_{\ell m n}=\frac{A_{\ell m }^{(\text{ref})}(\omega_{\ell m n})}{2 \omega_{\ell m n}^3}\left(\frac{dA^{(\rm inc)}_{\ell m}}{d\omega}\right)^{-1}_{\omega=\omega_{\ell mn}}\,.
\end{equation}

To compute the EFs we must evaluate $c_{\ell m}^{(\text{out})}(\omega_{\ell mn})$ and $(dc_{\ell m}^{(\text{in})}/d\omega)_{\omega = \omega_{\ell mn}}$, which can be found by solving the generalized Teukolsky equation as discussed above. We compute them using two independent approaches. The first approach, introduced in Ref.~\cite{Oshita:2021iyn}, is based on the representation of the solutions of Eq.~\eqref{eq:radial_Teukolsky} through the Heun functions of \textsc{Mathematica}, namely \texttt{HeunG}. 
The second approach, discussed in Appendix~\ref{appendix:STU}, is a generalization of the STU/MST method~\cite{Mano:1996vt,Suzuki:1999nn} and is based on a representation of the solutions of Eq.~\eqref{eq:radial_Teukolsky} in terms of hypergeometric functions, with the corresponding coefficients computed using the continued fraction method.
Unlike the original STU/MST method for the Kerr spacetime, this approach does not require a representation of the solutions in terms of Coulomb functions, because the background geometry is not asymptotically flat and the integration terminates at the cosmological horizon.

\begin{figure}
    \centering
    \includegraphics[width=\linewidth]{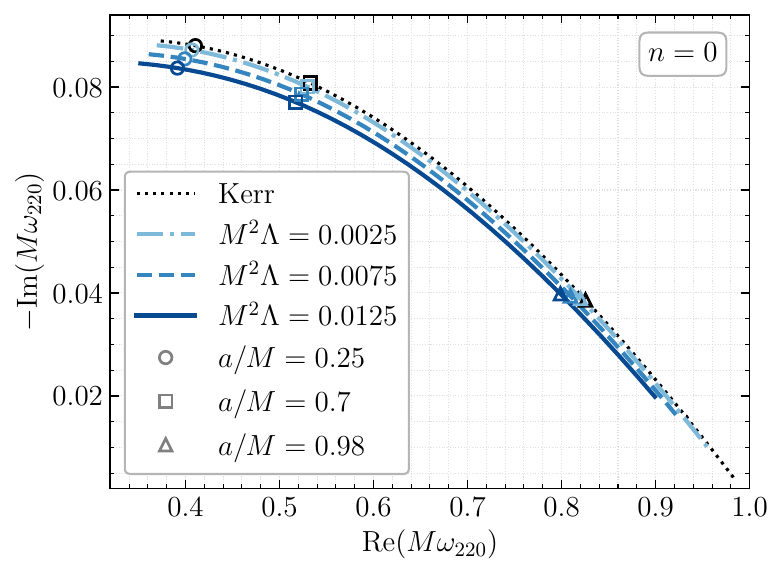}
    \caption{Fundamental KdS QNM frequencies with $(\ell,\,m,\,n)=(2,\,2,\,0)$ for selected values of $M^2{\rm \Lambda} = 0.0025,\,0.0075,\,0.0125$, denoted with increasingly darker colors and different linestyles. The dotted line is the corresponding Kerr QNM. Hollow circles, squares and triangles correspond to dimensionless spins $a/M=0.25,\,0.7$ and $0.98$, respectively.}
    \label{fig:systematic0}
\end{figure}

\begin{figure*}[t]
    \centering    
    \includegraphics[width=\linewidth]{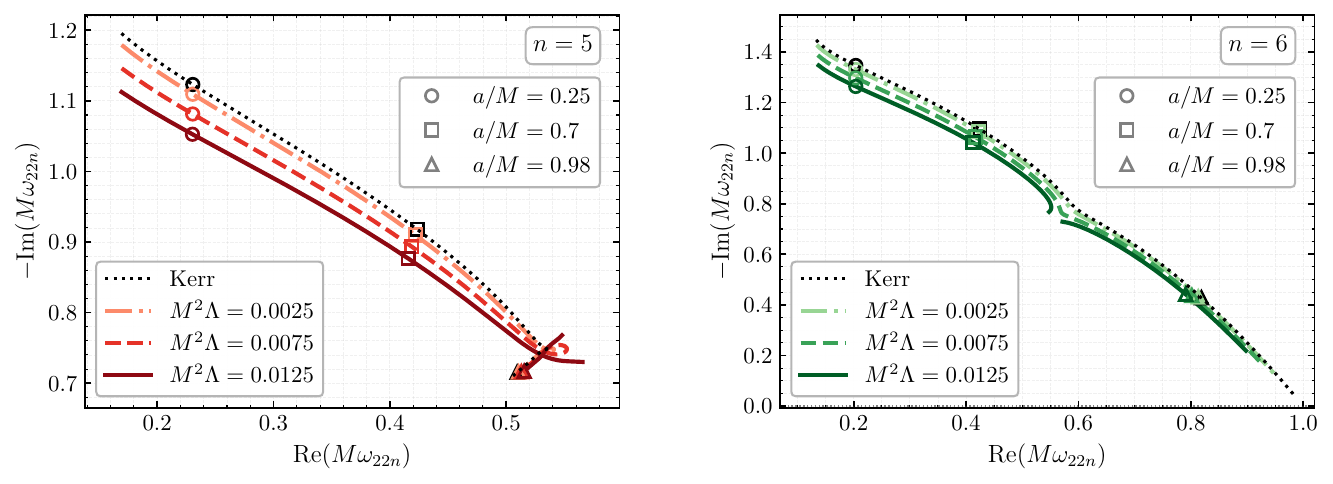}
    \caption{Same as Fig.~\ref{fig:systematic0}, but for the $n=5$ (left) and the $n=6$ (right) overtones with $(\ell,\,m)=(2,\,2)$. The reason for the discontinuity in the dark green line on the right panel is clarified in Fig.~\ref{fig:l2m2lambda004} below.}
    \label{fig:systematic1}
\end{figure*}

\begin{figure*}[t]
    \centering
    \includegraphics[width=\linewidth]{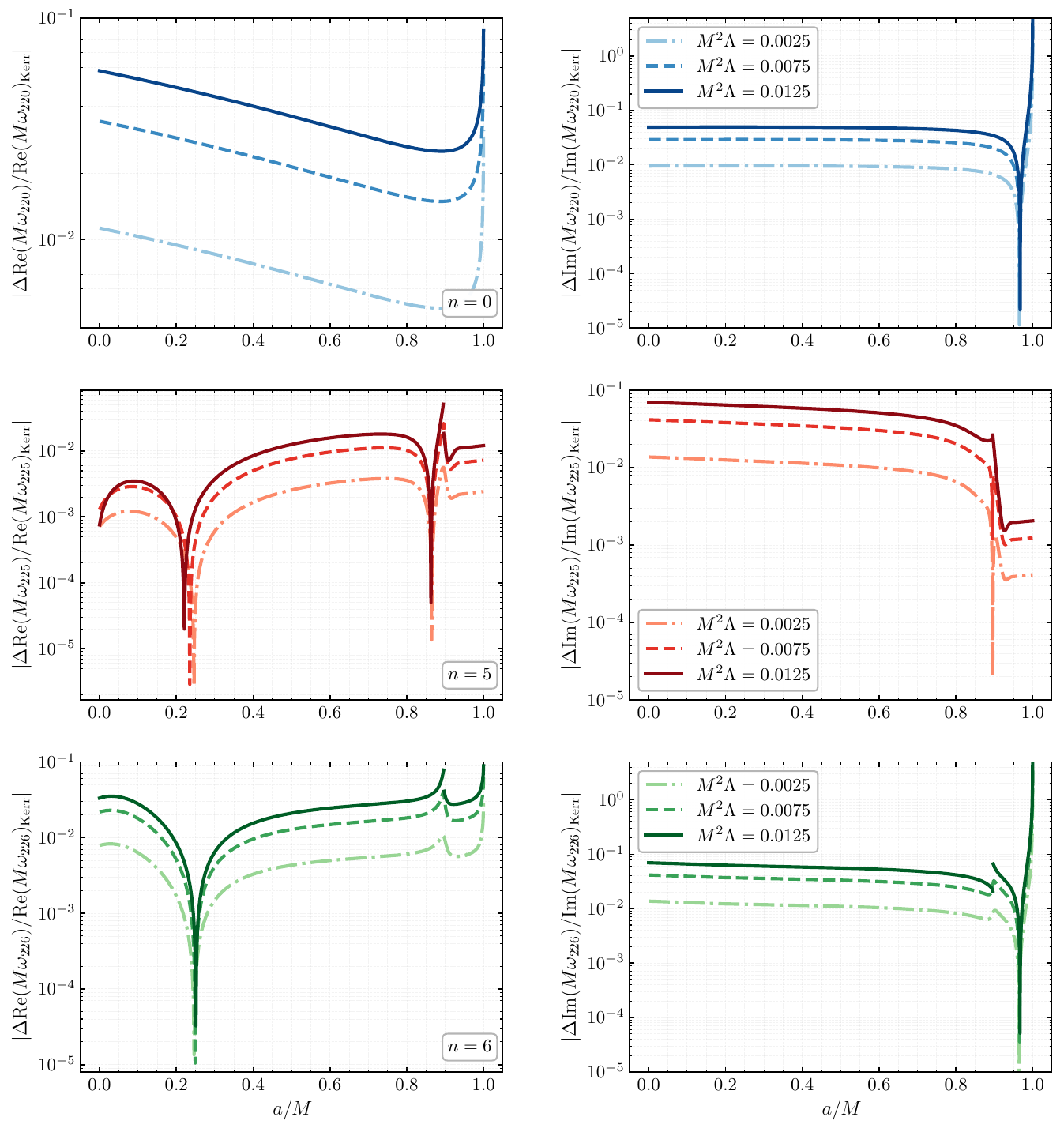}
    \caption{Relative difference in the real (left) and imaginary (right) QNM frequencies of KdS and Kerr spacetimes for the $(\ell,\,m)=(2,\,2)$ fundamental mode (top), the $n=5$ overtone (center) and the $n=6$ overtone (bottom).}
    \label{fig:systematic}
\end{figure*}

As a consistency check, we have compared the KdS EFs in the limit $M^2{\rm \Lambda} \to0$ with those computed in the Kerr spacetime. A similar comparison was performed in~\cite{Oshita:2021iyn} for $n=0,\,1,\,2$. Here we also consider the EFs of QNMs with $n=5,\,6$, that can be compared with the Kerr EFs recently computed in~\cite{Motohashi:2024fwt,motohashi_2024_12696857}.
We take the Kerr limit of the EFs following the formalism developed in~\cite{Oshita:2021iyn}. The procedure used to find the Kerr EFs from the KdS EFs is discussed in Appendix~\ref{appendix:EF_Kerr_limit}.

As pointed out in Ref.~\cite{Kubota:2025hjk}, the tortoise coordinate $r_\star$ used here and in Ref.~\cite{Oshita:2021iyn} has a different integration constant with respect to the (Kerr) tortoise coordinate used in~\cite{Motohashi:2024fwt,Kubota:2025hjk} and in previous work on EFs~\cite{Zhang:2013ksa}. As clarified in Table~I of~\cite{Kubota:2025hjk}, for $M^2 {\rm \Lambda}=0$ the EFs computed with the two definitions differ by a multiplicative factor 
\begin{equation}
\label{eq:EF_conversionFactor}
    F=(1-a^2/M^2)^{i\omega}\,.
\end{equation}
Taking into account this conversion factor, the KdS EFs converge to those of Kerr spacetime, provided in Refs.~\cite{Motohashi:2024fwt,motohashi_2024_12696857}, for $M^2 {\rm \Lambda }\to 0$
within $\lesssim1\%$ up to spin $a/M\sim 0.99$, and around $\sim2\%$ in the extremal regime $a/M>0.99$, where the factor~\eqref{eq:EF_conversionFactor} diverges. An exception is the $(\ell,\,m,\,n)=(2,\,2,\,5)$ mode in the extremal regime, for which our calculation is unstable (see Appendix~\ref{appendix:EF_Kerr_limit}). We remark that the two EPs of KdS studied in this work occur at spins $a/M<0.99$, where our EF calculation is reliable.

\section{The quasinormal modes of Kerr-de Sitter black holes}
\label{sec:KdS_QNM_spectrum}

We perform a systematic study of KdS QNMs to understand how the cosmological constant affects the values of the QNM frequencies. The two numerical approaches we used --- the one based on Heun functions and the one based on continued fractions --- are in agreement, with relative differences that are typically of order $10^{-15}$ (and at worst of order of $10^{-4}$ for small values of $M^2{\rm \Lambda} \sim 10^{-3}$ and near extremality).

In Figs.~\ref{fig:systematic0} and \ref{fig:systematic1}, we show the $(\ell,\,m)=(2,\,2)$ QNMs of KdS BHs with $n=0$, $5$ and $6$. %
We discuss the EP corresponding to the $n=5$ and $n=6$ overtones in the next section.
In Fig.~\ref{fig:systematic} we show in the left (right) panels the relative difference between the real (imaginary) part of the QNM frequencies of KdS and Kerr spacetimes, as functions of the BH spin, for the same QNMs displayed in Figs.~\ref{fig:systematic0} and~\ref{fig:systematic1}.
In general, the cosmological constant reduces the damping time of the modes.
As in Kerr spacetime, the $(\ell,\,m,\,n)=(2,\,2,\,5)$ mode remains a damped mode for all values of the spin, while other modes become zero-damping modes in the extremal limit (see~\cite{Yang:2012pj,Yang:2013uba} for a discussion of zero-damping modes).
We also note a discontinuity in the branch with $M^2{\rm \Lambda} = 0.0125$ of the $n=5$ and $n=6$ modes. This discontinuity is related to the EP between the two overtones corresponding to $M^2{\rm \Lambda} \sim0.0085$ and $a/M\sim0.9$~\cite{Oshita:2025ibu}, as we discuss below.

\subsection{Exceptional points and hysteresis}
\label{sec:exceptional_points}
It is well known that certain branches $\omega_{\ell mn}$ and $\omega_{\ell mn'}$ of the Kerr QNM spectrum exhibit {\it avoided crossings} in the complex plane: as the spin is increased at fixed $(\ell,\,m)$, the QNM frequencies corresponding to different overtones come very close to each other and repel (see e.g.~\cite{Onozawa:1996ux,Berti:2004um,Berti:2004md,Cook:2014cta}, and~\cite{Berti:2025hly} for a review).
When an additional parameter is introduced and tuned, the two QNM frequencies may become degenerate at some specific value of the parameter, leading to an EP.
The presence of EPs was found for Kerr BHs coupled to a massive scalar field~\cite{Cavalcante:2024kmy,Cavalcante:2025abr} and, more recently, for Kerr-Newman BHs coupled to a massless scalar field~\cite{Cavalcante:2026vgr}.
For KdS BHs, the avoided crossing between the QNM frequencies $\omega_{225}$ and $\omega_{226}$~\cite{Motohashi:2024fwt} becomes an EP at a nonzero value of the cosmological constant~\cite{Oshita:2025ibu}.

Here we extend the analysis of Ref.~\cite{Oshita:2025ibu} to identify other EPs in the range $2\le\ell\le6$, $|m| \leq \ell$, and $0\le n\le 7$. 
We find another EP in the KdS spectrum for $(\ell,\,m)=(3,\,1)$ -- more specifically, a spectral degeneracy between the $\omega_{315}$ and $\omega_{316}$ overtones.
We also confirm that the QNM spectrum $\omega_{\ell mn}(a/{\rm \Lambda})$ exhibits the hysteresis phenomenon near the EPs that was observed for Kerr BHs coupled to massive scalar fields~\cite{Cavalcante:2024swt,Cavalcante:2025abr, Cavalcante:2024kmy} and for Kerr-Newman BHs coupled to massless scalar fields~\cite{Cavalcante:2026vgr}.
In the following subsections we focus on our findings for QNMs with $(\ell,\,m)=(2,\,2)$ and $(\ell,\,m)=(3,\,1)$, respectively.

\subsection{$\ell=2\text{,} \; m=2$ exceptional point}
\label{sec:QNM_l2m2}
\begin{figure}[t]
    \centering
    \includegraphics[width=\linewidth]{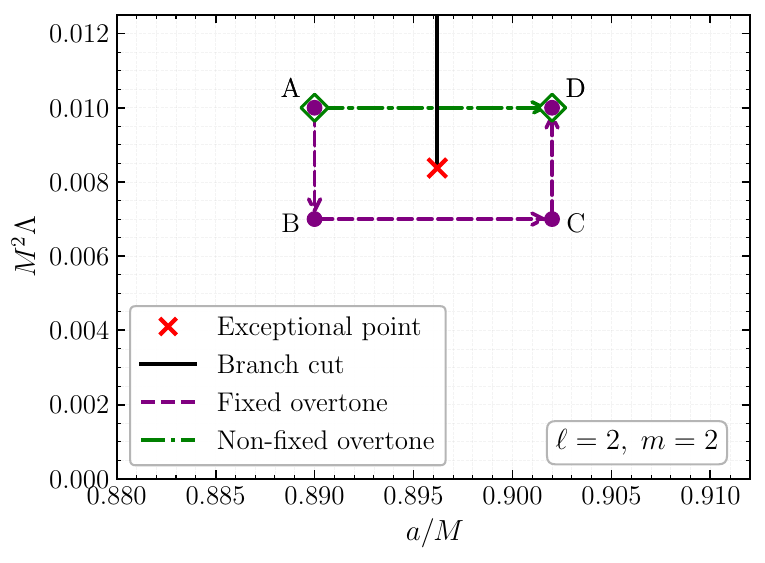}
    \caption{
      Representation of the branch cut (solid line) connecting the two separate Riemann sheets around the EP for $\omega_{225}$ and $\omega_{226}$ in the plane $(a/M,\,M^2{\rm \Lambda})$. The EP is the red cross at $(a/M)_{\rm EP}\simeq0.896$, $(M^2{\rm \Lambda})_{\rm EP}\simeq0.0085$. Starting from point A, point D can be reached following either the purple, dashed path (ABCD) that does not cross the branch cut, remaining in the same Riemann sheet of point A; or the green, dot-dashed path (AD) that crosses the branch cut, leading to a different Riemann sheet. At the final point D, the two paths lead to two different frequencies. The dash-dotted path leads to $M\omega_{225}({\rm D})|_{\rm dotdashed}=0.588868 - 0.729657 \,i$ and $M\omega_{226}({\rm D})|_{\rm dotdashed} =0.541571 - 0.75595\,i$, while the dashed yields $M\omega_{225}({\rm D})|_{\rm dashed} =0.541571 - 0.75595\, i$ and $M\omega_{226}({\rm D})|_{\rm dashed}=0.589137 - 0.730582\, i$.}
    \label{fig:parameterPlane_l2m2}
\end{figure}

\begin{figure*}[t]
    \centering
    \includegraphics[width=\linewidth]{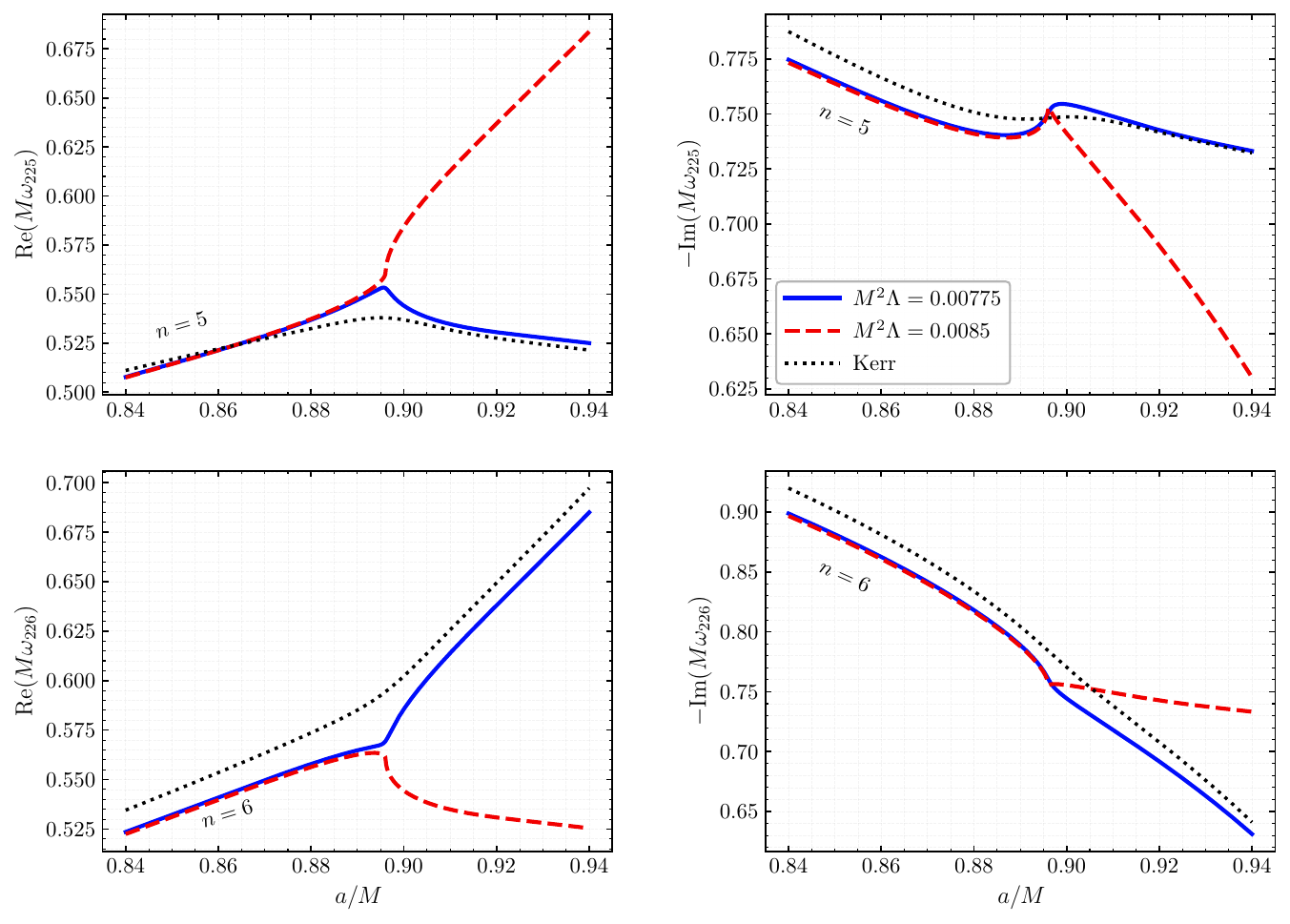}
    \caption{Real (left) and imaginary (right) parts of the $(\ell,\,m)=(2,\,2)$ KdS QNM frequencies along curves with constant $M^2{\rm \Lambda}$, as the spin changes over a range that includes the EP. Solid (dashed) lines show the frequencies with cosmological constant just below (above) the EP value $(M^2{\rm \Lambda})_{\rm EP}$. The curves on the top (bottom) panels start close to the $n=5$ ($n=6$) Kerr overtone for spins below $(a/M)_{\rm EP}$, and change overtone number (i.e., they get closer to the ``other'' Kerr QNM) as they cross the EP. For reference, dotted lines in the top (bottom) panel show the Kerr QNMs with $n=5$ ($n=6$).}
    \label{fig:ReImVSspin_separation_l2m2}
\end{figure*}

\begin{figure*}[t]
    \centering
    \includegraphics[width=\linewidth]{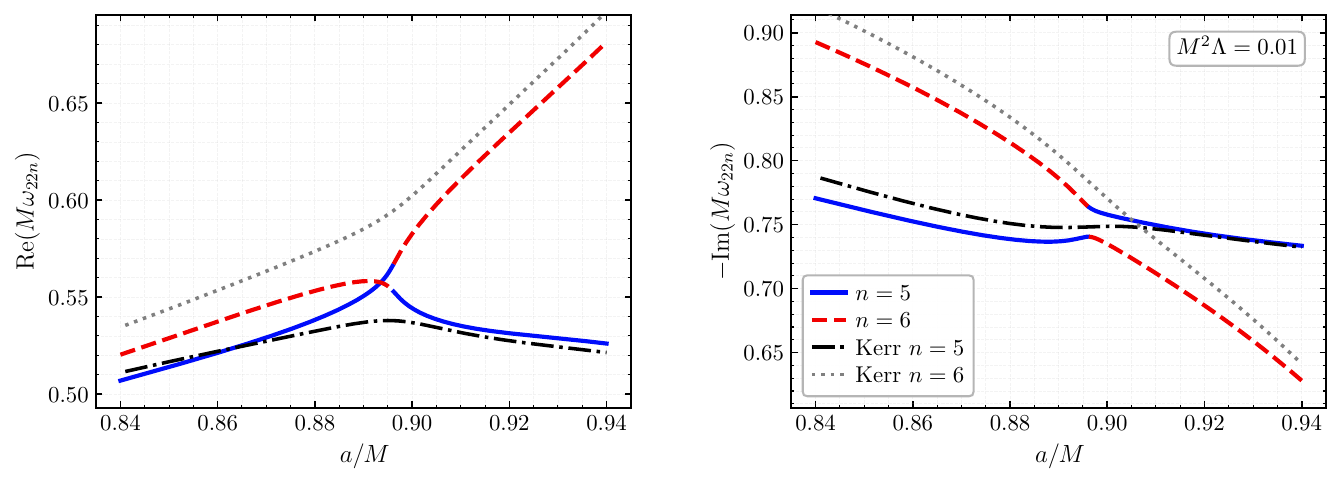}
    \caption{Real (left) and imaginary (right) parts of the $(\ell,\,m)=(2,\,2)$ QNM frequencies with fixed overtone numbers (blue for $n=5$, red for $n=6$). The curves are discontinuous as they cross the branch cut at $a/M=(a/M)_{\rm EP}$. For reference, the black lines show the Kerr QNMs with $n=5$ (dash-dotted) and $n=6$ (dotted).}
    \label{fig:l2m2lambda004}
\end{figure*}

We identify the EP parameters $(a/M)_{\rm EP}$ and $(M^2{\rm \Lambda})_{\rm EP}$ by computing the frequencies in a range close to the Kerr avoided crossing between the overtones $n=5$ and $n=6$, as the cosmological constant in increased. We find that the real and imaginary parts of the QNM frequencies of the two overtones are the same up to the eighth decimal place for $a/M\simeq0.896$ and $M^2{\rm \Lambda}\simeq0.0085$, in agreement with Ref.~\cite{Oshita:2025ibu}. 

In practice, in order to find the modes for general values of $a/M$ and $M^2{\rm \Lambda}$, we need to follow some curve in the $(a/M,\,M^2{\rm \Lambda})$ plane.
In both of our numerical approaches we use a Newton-Raphson algorithm where, at each step along the curve, the trial value is close to the correct QNM frequency. Remarkably, we find that different curves around the EP approaching the same value of $(a/M,\,M^2{\rm \Lambda})$ lead to different QNM frequencies, corresponding to different overtones. This is the hysteresis phenomenon discussed in Refs.~\cite{Cavalcante:2024kmy,Yang:2025dbn,Cavalcante:2026vgr}.

The hysteresis mechanism is clarified in Fig.~\ref{fig:parameterPlane_l2m2}, where we show different curves in the $(a/M,\,M^2{\rm \Lambda})$ plane. The red cross denotes the EP. 
If we start from the $n=5$ mode at point $A$, we can find the complex mode frequency at point $D$ following different paths. If we follow the dashed path going from $A$ to $B$ at constant $a$, from $B$ to $C$ at constant $M^2{\rm \Lambda}$, and from $C$ to $D$ at constant $a$, the mode solution at the point $D$ is close (within a few percent) to the value of the $n=5$ Kerr mode for the same $a/M$. If instead we go directly from $A$ to $D$ at constant $M^2{\rm \Lambda}$ along the dash-dotted path, we find a {\it different} value of the mode frequency, which turns out to be close (within a few percent) to the $n=6$ Kerr QNM frequency for the same $a/M$. If we now repeat the same procedure but we use as our initial guess the $n=6$ mode at point $A$ rather than the $n=5$ mode, the QNM frequencies that we find at point $D$ are swapped.

These results can be interpreted by treating the $(a/M,\,M^2{\rm \Lambda})$ plane as a complex plane, and $\omega_{\ell mn}$ as an analytic, multi-valued function on this plane. A closed path containing the EP changes leads to a different Riemann as we cross the branch cut. Formally, any line starting at the EP could be chosen as a branch cut. In our case, we choose to associate the two Riemann sheets to the $n=5$ and $n=6$ QNMs, respectively. As we will see, this choice corresponds to the constant-spin branch cut with $M^2{\rm \Lambda}>(M^2{\rm \Lambda})_{\rm EP}$ shown in the figure.

To further clarify the behavior of the QNM frequency near the EP,
in Fig.~\ref{fig:ReImVSspin_separation_l2m2} we show the mode frequencies computed following curves at two constant values of $M^2{\rm \Lambda}$, $M^2{\rm \Lambda}=0.00775$ (solid blue) and $M^2{\rm \Lambda}=0.0085$ (dashed red). We start at $a/M=0.84$ and increase the spin, using the mode frequency at each step as a trial value for the following step.
The top panels refer to the $n=5$ overtone, and the bottom panels to $n=6$.
For reference, we also show the Kerr QNMs ($M^2{\rm \Lambda}=0$, dotted black).
While the KdS QNM frequencies are initially close to the Kerr values, the curve with $M^2{\rm \Lambda}>(M^2{\rm \Lambda})_{\rm EP}$ departs from the Kerr QNMs after crossing the EP. Remarkably, for $a/M>(a/M)_{\rm EP}$, the dashed KdS modes in the top (bottom) panels are close to the Kerr modes in the bottom (top) panel. This happens because the dashed curves in Fig.~\ref{fig:ReImVSspin_separation_l2m2} swap their overtone numbers when crossing the EP -- i.e., for $M^2{\rm \Lambda}>(M^2{\rm \Lambda})_{\rm EP}$, $a/M>(a/M)_{\rm EP}$ the KdS QNMs in the top panels correspond to $n=6$, and those in the bottom panels to $n=5$. However, we did not find a rigorous way to characterize the overtone numbers of these curves. In particular, we find that the radial profiles of the $n=5$ and $n=6$ eigenfunctions are very similar near the EP,
so the overtone number labeling is somewhat arbitrary. It is then natural to label KdS modes by their proximity to the Kerr modes, as in the discussion above. With this choice, the overtone number (and the Riemann sheet) changes along the curve with constant $M^2{\rm \Lambda}$, while it remains the same along the curve with constant $a/M$.
This can be seen in Fig.~\ref{fig:l2m2lambda004}, where we plot the real (left) and imaginary (right) parts of the QNM frequencies for a given overtone number ($n=5$ in blue, and $n=6$ in red) along the path with $M^2{\rm \Lambda}=0.01$ as we increase $a/M$ around the EP value -- i.e., the dash-dotted green path of Fig.~\ref{fig:ReImVSspin_separation_l2m2}. In this case, since the overtone number is fixed, the modes have a discontinuity as they cross the branch cut.

\subsection{$\ell=3\text{,} \; m=1$ exceptional point}
\label{sec:QNM_l3m1}

\begin{figure}[t]
    \centering
    \includegraphics[width=1.005\linewidth]{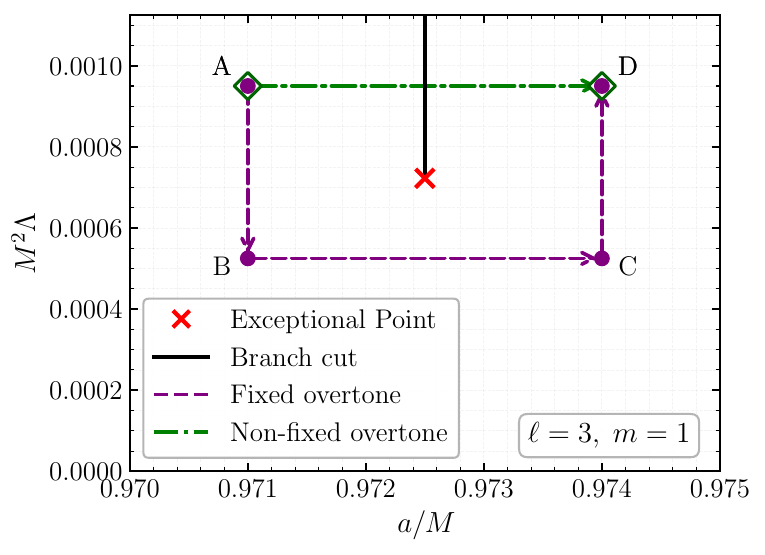}
    \caption{Same as Fig.~\ref{fig:parameterPlane_l2m2}, but for $(\ell,\,m)=(3,\,1)$.}
    \label{fig:parameterPlane_l3m1}
\end{figure}

\begin{figure*}[t]
    \centering
    \includegraphics[width=\linewidth]{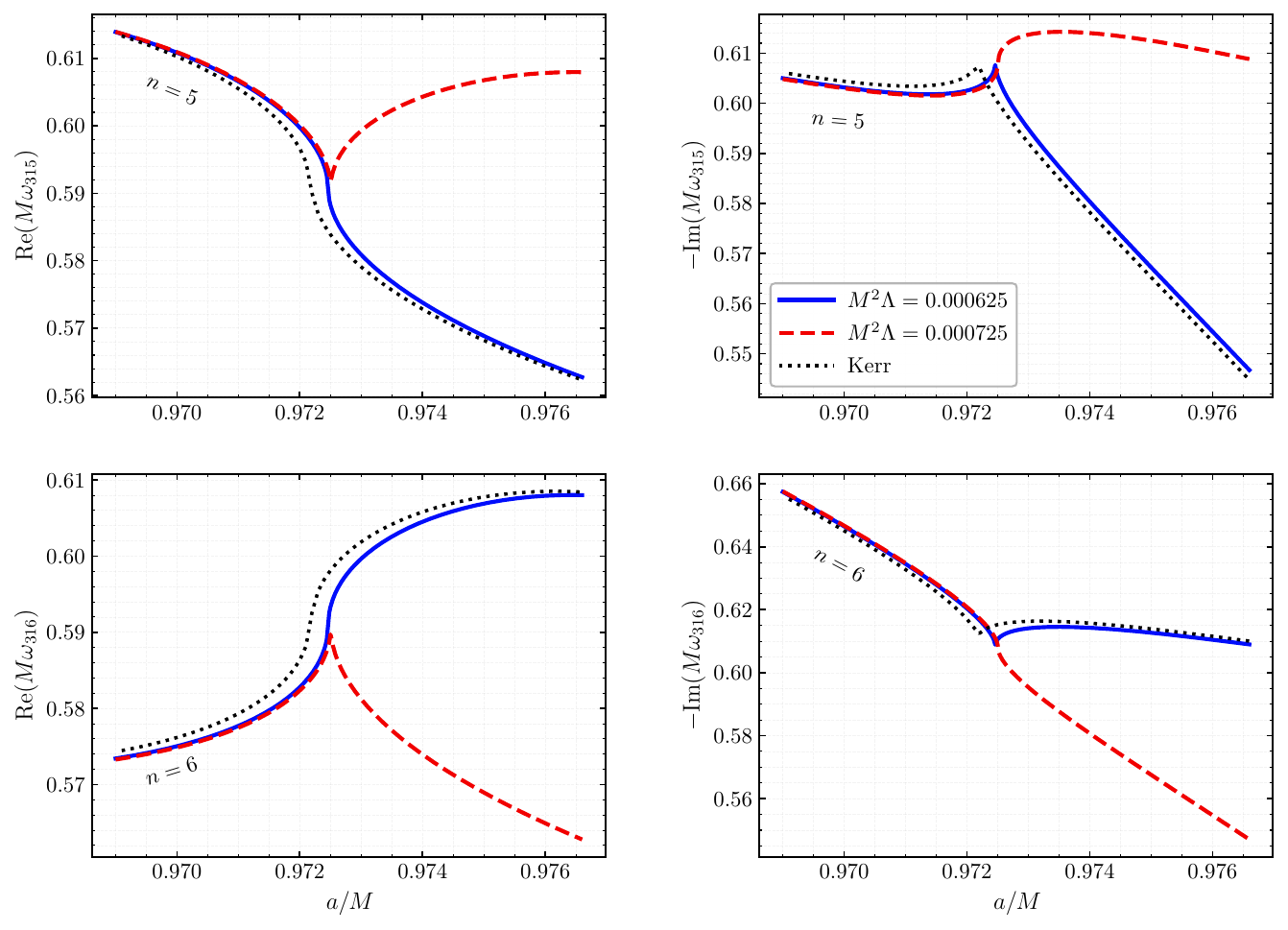}
    \caption{Same as Fig.~\ref{fig:ReImVSspin_separation_l2m2}, but for the $\omega_{315}$ and $\omega_{316}$ QNMs.}
    \label{fig:ReImVSspin_separation_l3m1}
\end{figure*}

\begin{figure*}[t]
    \centering
    \includegraphics[width=\linewidth]{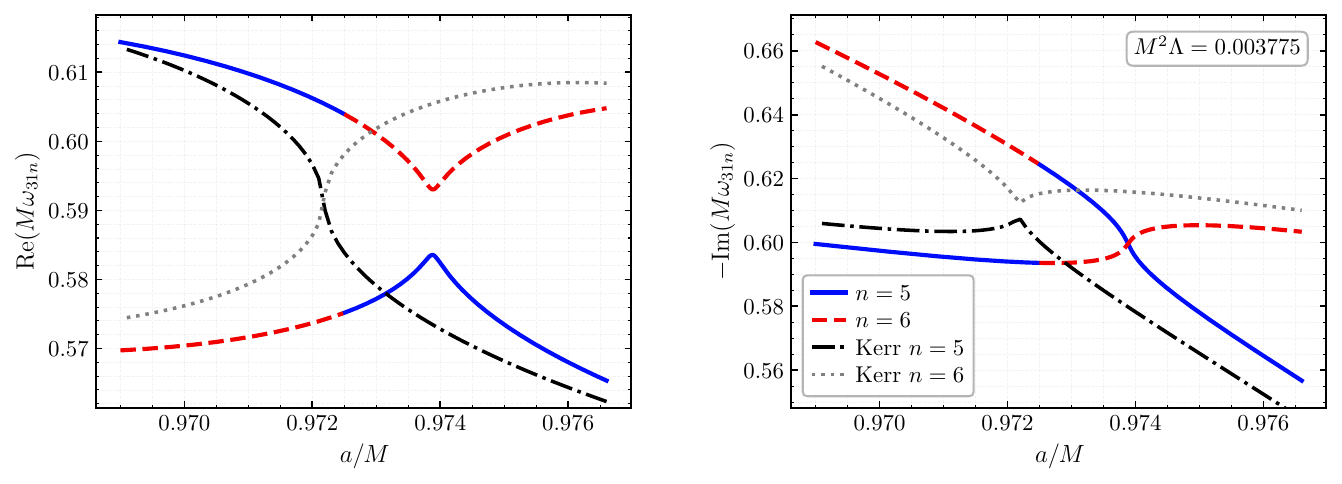}
    \caption{Same as Fig.~\ref{fig:l2m2lambda004}, but for the $\omega_{315}$ and $\omega_{316}$ QNMs.}
    \label{fig:l3m1lambda000151}
\end{figure*}

For $(\ell,\,m)=(3,\,1)$ we find an EP between the modes $\omega_{315}$ and $\omega_{316}$, located at $(M^2{\rm \Lambda})_{\rm EP}\simeq0.000725$ and $(a/M)_{\rm EP}\simeq0.9725$ (this EP corresponds to a Kerr avoided crossing identified in the past~\cite{Onozawa:1996ux,Berti:2004um,Berti:2004md,Cook:2014cta} and discussed in Ref.~\cite{Motohashi:2024fwt,Kubota:2025hjk}). 
When compared with the EP involving the $\omega_{225}$ and $\omega_{226}$ frequencies discussed above, the cosmological constant is about one order of magnitude smaller, and the spin closer to extremality.

In Fig.~\ref{fig:parameterPlane_l3m1} we perform the same analysis that we carried out in Sec.~\ref{sec:QNM_l2m2}, and we show different paths in the $(a/M, \,M^2{\rm \Lambda})$ plane around the $(\ell,\,m)=(3,\,1)$ EP (denoted by a red cross). When we follow the dash-dotted path from A to D, we find a mode different from the mode that we would find by following the dashed path: the dash-dotted path crosses the branch cut, and the overtone number (that we label by its proximity with the corresponding Kerr QNM, as in Sec.~\ref{sec:QNM_l2m2}) changes.

In Fig.~\ref{fig:ReImVSspin_separation_l3m1} -- the $(\ell,\,m)=(3,\,1)$ analog of Fig.~\ref{fig:ReImVSspin_separation_l2m2} -- we show modes computed by increasing the spin along curves with constant $M^2{\rm \Lambda}$, while in Fig.~\ref{fig:l3m1lambda000151} -- the analog of Fig.~\ref{fig:l2m2lambda004} -- we fix the overtone number. The comparison with the Kerr modes (dotted curve) shows that QNMs with different overtone numbers swap places in the first case, while the curves are discontinuous in the second case.

\section{Excitation Factors}
\label{sec:excitation_factors}

The EFs $E_{lmn}$ defined in Sec.~\ref{subsec:efKdS} quantify how the various modes can be excited by a generic source. In Kerr spacetime, near avoided crossings, the mode repulsion leads to a large enhancement of the amplitude of the EFs~\cite{Yang:2025dbn,Motohashi:2024fwt,Kubota:2025hjk}.
In KdS, near an EP, the enhancement of the EFs is even larger~\cite{Oshita:2025ibu}.

\subsection{Comparison of different approaches}

\begin{figure*}[t]
    \centering
    \includegraphics[width=\linewidth]{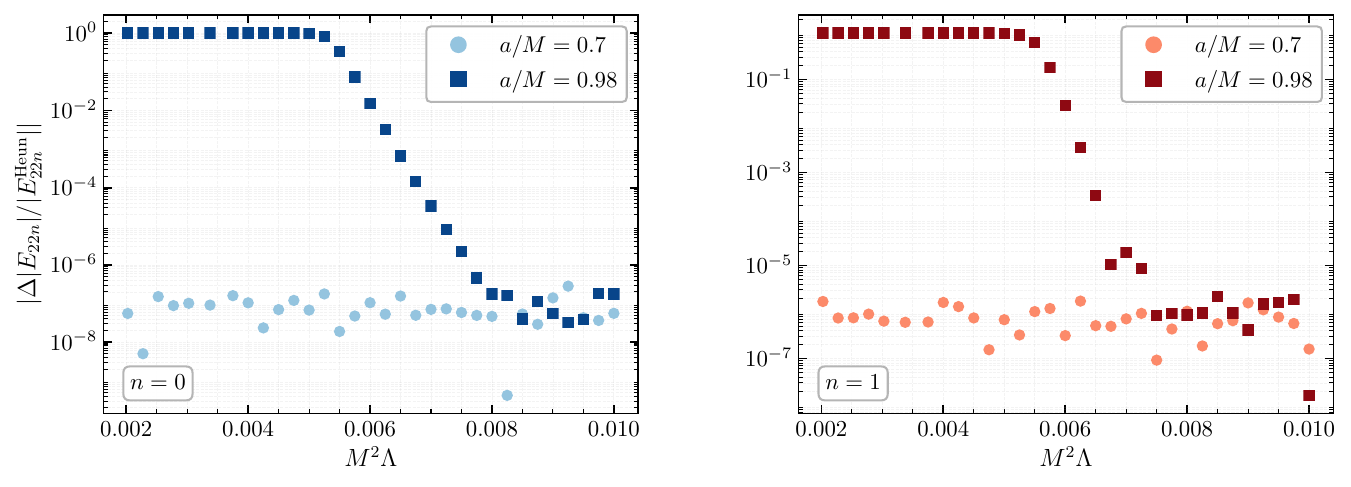}
    \includegraphics[width=0.5\linewidth]{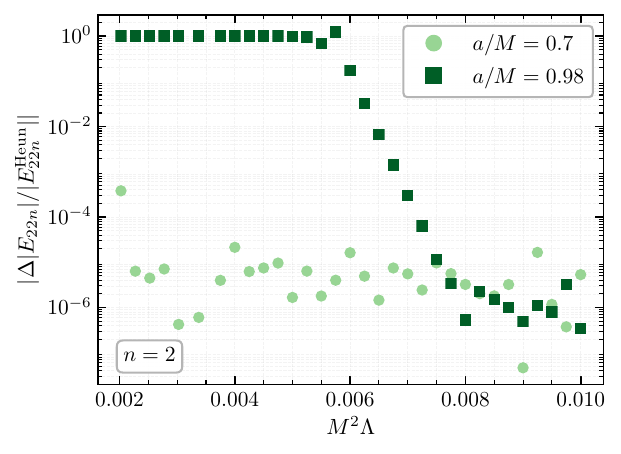}
    \caption{Relative difference between the absolute value of the EFs computed with the two different approaches (Heun functions and STU, see Appendix~\ref{appendix:STU}) as a function of $M^2{\rm \Lambda}$, for $(\ell,\,m)=(2,\,2)$ and overtones $n=0,1,2$. Here, $E^{\rm Heun}_{\ell m}$ denotes the EFs computed using the built-in \textsc{Mathematica} Heun functions \texttt{HeunG}.}
    \label{fig:EF_n012_same_nuSeries}
\end{figure*}

\begin{figure*}[t]
    \centering
    \includegraphics[width=\linewidth]{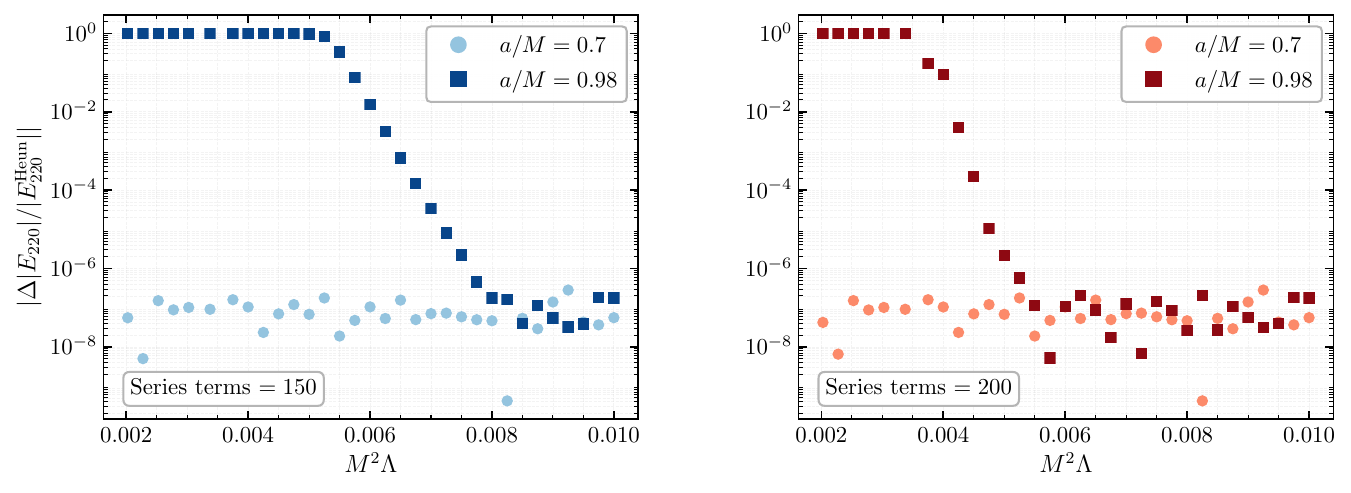}\\
    \includegraphics[width=0.5\linewidth]{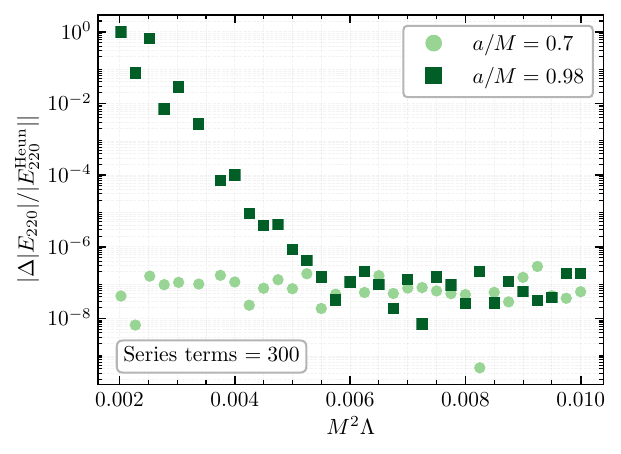}
    \caption{Relative difference between the absolute value of the EFs computed with the two different approaches (Heun functions and STU) as a function of $M^2{\rm \Lambda}$, for $(\ell,\,m,\,n)=(2,\,2,\,0)$. A comparison of the different panels shows that increasing the number of terms in the series of hypergeometric functions~\eqref{eq:hypergeometric_heun_sol} improves the agreement at small $M^2{\rm \Lambda}$ for $a/M=0.98$.}
    \label{fig:EF_n0_different_nuSeries}
\end{figure*}

\begin{figure*}[t]
    \centering
    \includegraphics[width=\linewidth]{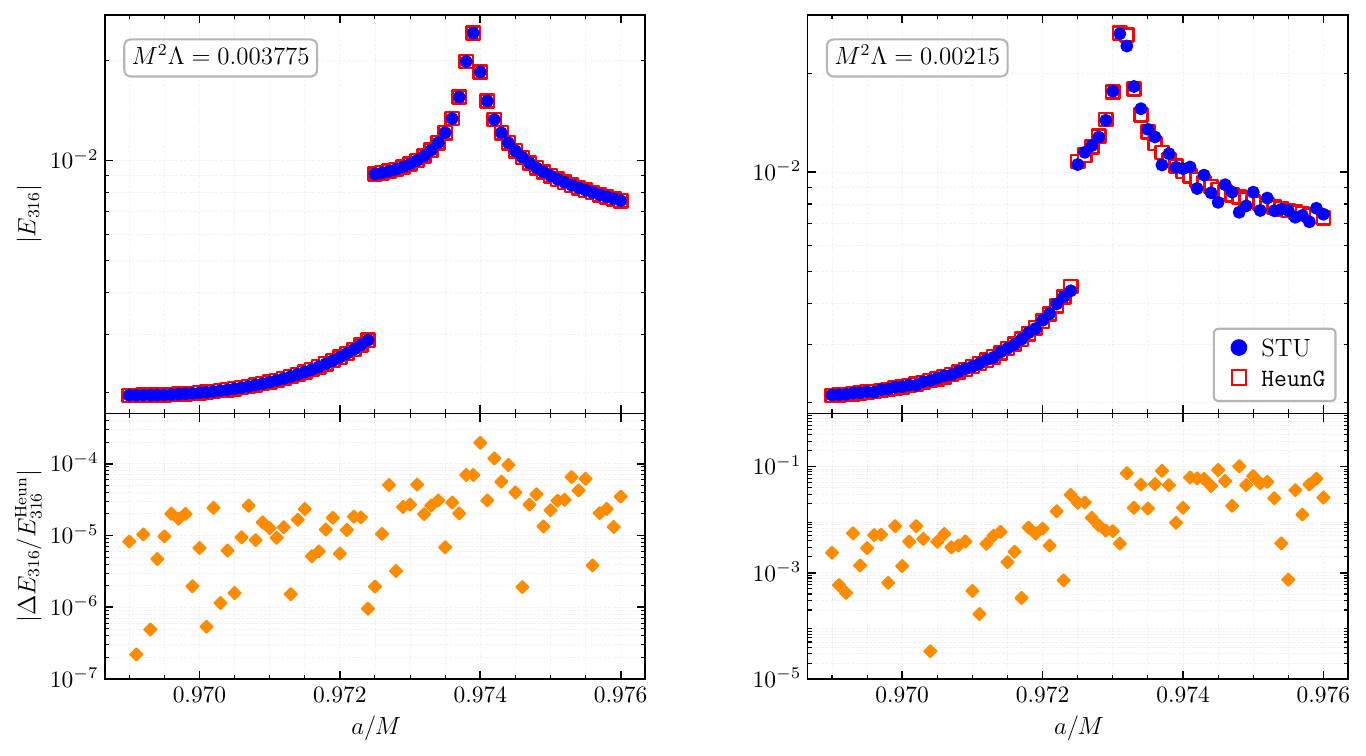}
    \caption{Comparison between the absolute value of the $E_{316}$ EF computed using the STU method of Appendix~\ref{appendix:STU} (blue circles) and the \textsc{Mathematica} Heun functions method of Ref.~\cite{Oshita:2021iyn} (red squares). In general the two methods are in very good agreement ($|\Delta E_{\ell mn}/E^{\rm Heun}_{\ell mn}|<10^{-4}$), but for $M^2{\rm \Lambda} \lesssim 3\times 10^{-3}$ (right panel) and for large values of $a/M$ the STU calculation has some numerical noise that is not present in the Heun function calculation.}
    \label{fig:STUvsHEUN}
\end{figure*}

Since the numerical computation of the EFs in KdS spacetime is technically challenging~\cite{Oshita:2021iyn}, here we compute them using two independent approaches: the Heun function approach developed in Ref.~\cite{Oshita:2021iyn}, and the STU continued fraction approach based on hypergeometric functions discussed in Appendix~\ref{appendix:STU}.

In Fig.~\ref{fig:EF_n012_same_nuSeries} we show the relative difference between the absolute values of the EFs computed with the two methods as a function of $M^2{\rm \Lambda}$, for $(\ell,\,m)=(2,\,2)$ and overtones $n=0,1,2$, fixing the value of the number of terms in the series of hypergeometric functions of Eq.~\eqref{eq:hypergeometric_heun_sol} to 150. In the non-extremal case $a/M=0.7$, the agreement between the two methods ranges between $\sim 10^{-8}$ and $\sim 10^{-4}$, with generally better agreement for lower overtone numbers. For the larger spin values (as illustrated in the figure for $a/M=0.98$) the agreement becomes much worse as $M^2{\rm \Lambda}\to 0$. This can be explained by the fact that the Heun function around the cosmological horizon ($z=1$), written in terms of the hypergeometric series in the STU method, has a convergence radius of $\text{min}(1,\,z_a)$, where $z_a\to 1$ when $M^2{\rm \Lambda}$ is small and $a/M\to1$; conversely, the convergence radius of the solution around the event horizon ($z=0$) is always 1. To determine the values of $c^{\rm(out/in)}_{\ell m}$ in Eqs.~\eqref{eq:RinBH}, we need to compute the Wronskians of the ingoing solution at the BH event horizon and the ingoing/outgoing solutions at the cosmological horizon~\cite{Hatsuda:2020sbn,Motohashi:2021zyv}. When the convergence radius for the solutions around the cosmological horizon shrinks, the overlap region where the series solutions around the two horizons are both convergent is pushed away from $z=0$ and towards $z=1$, so an accurate series solution at the event horizon must include more terms.
In Fig.~\ref{fig:EF_n0_different_nuSeries} we show that, indeed, increasing the number of terms in the series improves the agreement between the two methods for small values of $M^2{\rm \Lambda}$, even when $a/M=0.98$. Note that the method that uses \textsc{Mathematica}'s built-in \texttt{HeunG} function is not affected by this issue, because the built-in function can be analytically continued outside of its convergence radius~\cite{Motohashi:2021zyv}.

The EFs of the $(\ell,\,m)=(2,\,2)$ EP are computed for $M^2{\rm \Lambda}\sim 0.008$ and $a/M\sim0.9$, where the two approaches are always in good agreement ($|\Delta E_{\ell mn}/E^{\rm Heun}_{\ell mn}|<10^{-4}$) when $300$ terms are included in the series. 
However, the EFs of the $(\ell,\,m)=(3,\,1)$ EP must be computed in a region where $M^2{\rm \Lambda}\sim10^{-4}$ and $a/M\gtrsim 0.97$. In order to ensure agreement between the two methods we should include a very large number of terms in the STU series, which is computationally time-consuming. 
In general, since the two methods agree with very good accuracy, they validate each other.
In Fig.~\ref{fig:STUvsHEUN} we compare $E_{316}$ computed using the STU method (blue circles) and the \texttt{HeunG} functions approach (red squares). The discontinuity in the curves correspond to the branch cut. When $M^2{\rm \Lambda}=0.00215$ (right panel) the results obtained with the \texttt{HeunG} method vary continuously for spins up to $a/M\sim0.98$, while the results from the STU method show some numerical noise for spins $a/M\gtrsim0.9725$, even when $450$ terms are included in the series. In general, the \texttt{HeunG} calculation is our method of choice for spins up to $a/M\sim0.98$ and for $M^2{\rm \Lambda}\lesssim 2.5\times10^{-3}$.

As a further check of our KdS EF calculation, we have checked that in the limit $M^2{\rm \Lambda}\to0$ we reproduce the online catalog of Kerr EFs provided by Motohashi~\cite{Motohashi:2024fwt,motohashi_2024_12696857} with a precision of $\sim 1\%$, as we discussed in Sec.~\ref{subsec:efKdS} and (in some more detail) in Appendix~\ref{appendix:EF_Kerr_limit}.

\subsection{Excitation factors near exceptional points}

\begin{figure*}[t]
  \centering
  \includegraphics[width=\linewidth]{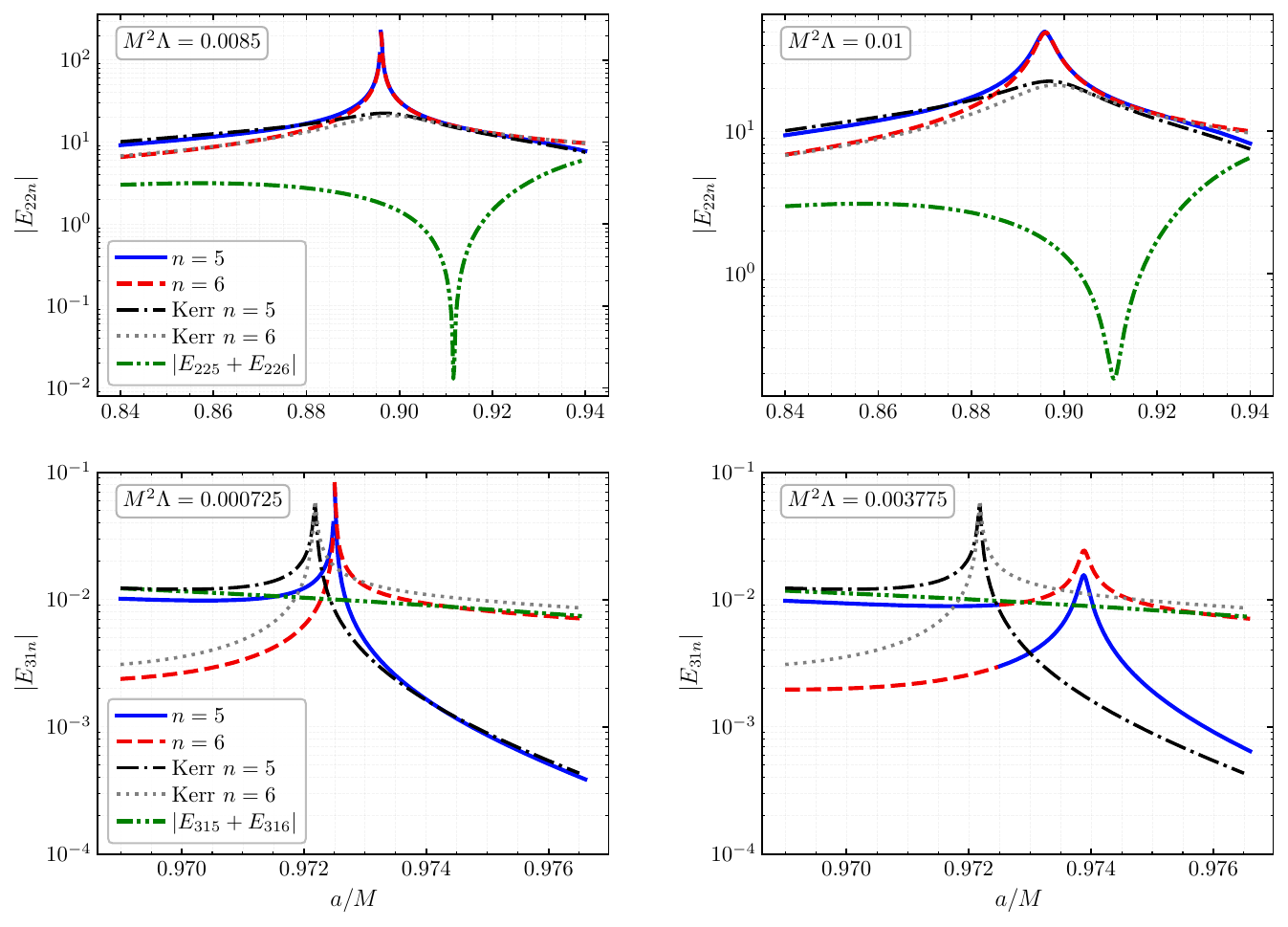}
  \caption{Absolute value of the EFs for fixed overtone numbers $n=5$ and $n=6$ of modes $(\ell,\,m)=(2,\,2)$ (top panels) and $(\ell,\,m)=(3,\,1)$ (bottom panels), for values of $M^2{\rm \Lambda}$ at the corresponding EP (left panels) and $ M^2{\rm \Lambda}>(M^2{\rm \Lambda})_{\rm EP}$ (right panels). Black curves refer to the Kerr EFs for the same modes, taken from the catalog~\cite{Motohashi:2024fwt,motohashi_2024_12696857} and multiplied by the factor $F$ of Eq.~\eqref{eq:EF_conversionFactor}. Green dash-dot-dotted lines show $|E_{225}+E_{226}|$ (top panels) and $|E_{315}+E_{316}|$ (bottom panels).}
  \label{fig:EF}
\end{figure*}
  
In Fig.~\ref{fig:EF} we show the KdS EFs for $(\ell,\,m)=(2,\,2)$ (top panels) and $(\ell,\,m)=(3,\,1)$ (bottom panels) as functions of $a/M$, for selected constant values of $M^2{\rm \Lambda}$. In the left panels we set $M^2{\rm \Lambda}\simeq (M^2{\rm \Lambda})_{\rm EP}$, while in the right panels we have $M^2{\rm \Lambda}>(M^2{\rm \Lambda})_{\rm EP}$.
Blue (red) curves correspond to QNMs with $n=5$ ($n=6$). For $M^2{\rm \Lambda}>(M^2{\rm \Lambda})_{\rm EP}$, the EF curves are discontinuous as we cross the branch cut, just like the corresponding QNM frequencies (see Figs.~\ref{fig:l2m2lambda004} and~\ref{fig:l3m1lambda000151}). This is clearly visible for $(\ell,\,m)=(3,\,1)$, while for $(\ell,\,m)=(2,\,2)$ the discontinuity is very close to the peak of the amplitude of the EFs, and more resolution would be required to see it more clearly.

We confirm the enhancement of the EFs near the $(\ell,\,m)=(2,\,2)$ EP found in Ref.~\cite{Oshita:2025ibu}, and we find a similar behavior near the $(\ell,\,m)=(3,\,1)$ EP. As shown in Fig.~\ref{fig:EF},  near the EP there is an enhancement in the absolute value of the individual $E_{\ell mn}$'s with respect to their Kerr values for both $(\ell,\,m)=(2,\,2)$ and $(\ell,\,m)=(3,\,1)$.
When $M^2{\rm \Lambda} > (M^2{\rm \Lambda}) _{\rm EP}$, the (smaller) peak in the KdS EF amplitude as a function of $a/M$ remains close to its Kerr location for $(\ell,\,m)=(2,\,2)$, while it shifts to larger values of $a/M$ for $(\ell,\,m)=(3,\,1)$. As noted in~\cite{Yang:2025dbn}, this peak always corresponds to the value of the spin for which the $n=5$ and $n=6$ QNM frequencies are closest.

As discussed in Ref.~\cite{Oshita:2025ibu}, the EFs with $n=5,\,6$ have nearly opposite phases near the $(\ell,\,m)=(2,\,2)$ EP, therefore $|E_{225}+E_{226}|\ll |E_{225}|$ and we do not expect the EF enhancement to lead to enhanced gravitational-wave emission near the EPs. We find the same behavior near the $(\ell,\,m)=(3,\,1)$ EP, i.e., $|E_{315}+E_{316}|\ll |E_{315}|$. This is clear from the green dash-dot-dotted curves in Fig.~\ref{fig:EF}.

We remark that the black curves in Fig.~\ref{fig:EF} show the Kerr EFs computed in Ref.~\cite{Motohashi:2024fwt,motohashi_2024_12696857} normalized by $F/\omega^2$, where the factor $F$ (see Eq.~\eqref{eq:EF_conversionFactor} and the discussion above) is needed to match the different definition of the tortoise coordinate in~\cite{Motohashi:2024fwt,motohashi_2024_12696857} (see Refs.~\cite{Oshita:2024wgt,Kubota:2025hjk} for discussions), and the factor $\omega^2$ is needed to match the different definitions of the EFs in Eq.~\eqref{eq:excitation_factor} and in~\cite{Berti:2006wq,Zhang:2013ksa,Motohashi:2024fwt,Motohashi:2026mbn}.

\section{Conclusions}\label{sec:concl}

We have computed the QNMs and EFs of KdS BHs using two different techniques: a Heun function representation and an STU/MST expansion of the solution in terms of hypergeometric functions.

We have found and studied two EPs (although the presence of EPs seems to be a generic feature, and it is very well possible that additional EPs are present in the spectrum). For one of these EPs, corresponding to a crossing of the KdS QNM frequencies $\omega_{225}$ and $\omega_{226}$ that occurs when $M^2{\rm \Lambda}\simeq0.0085$ and $a/M\simeq0.896$, we have reproduced and extended the results of Ref.~\cite{Oshita:2025ibu}. The other EP is between the Kerr QNM frequencies $\omega_{315}$ and $\omega_{316}$ located at $(M^2{\rm \Lambda})_{\rm EP}\simeq0.000725$ and $(a/M)_{\rm EP}\simeq0.9725$, corresponding to the Kerr avoided crossing found previously in the literature~\cite{Onozawa:1996ux,Berti:2004um,Berti:2004md,Cook:2014cta} and discussed in Refs.~\cite{Motohashi:2024fwt,Kubota:2025hjk}. Near these EPs, the QNM frequencies exhibit a ``hysteresis phenomenon'' analogous to what was previously found for Kerr BHs perturbed by massive fields~\cite{Cavalcante:2024swt,Cavalcante:2024kmy,Cavalcante:2025abr} and for charged Kerr-Newman BHs~\cite{Cavalcante:2026vgr}. The absolute value of the EFs is significantly enhanced, but the EFs acquire nearly opposite phases, suggesting that their combined effect may not dramatically affect the resulting time-domain waveforms (see e.g.~\cite{Motohashi:2024fwt,Oshita:2025ibu,Yang:2025dbn,PanossoMacedo:2025xnf,Imafuku:2026rpn} for discussions).

By extrapolating the KdS EFs to small values of $M^2 {\rm \Lambda}$ we find generally good agreement with previous calculations in the Kerr limit~\cite{Motohashi:2024fwt,motohashi_2024_12696857}, but the calculation becomes numerically challenging and time consuming. When $M^2 {\rm \Lambda}\lesssim 3\times 10^{-3}$ the results obtained with the Heun function calculation seem to be numerically stable for spins up to $a/M\sim0.98$, while the results from the STU method show some numerical noise for spins $a/M\gtrsim 0.97$, even when a large number of terms ($\sim 450$) is included in the hypergeometric series expansion. 

Recent work~\cite{Arnaudo:2025uos,Arnaudo:2025kit,Arnaudo:2026tcy} has shown that Green’s functions of nonrotating, asymptotically de Sitter BH spacetimes can be expressed as a convergent mode sum everywhere in spacetime: at late times this sum involves QNMs, while at early times it involves Matsubara (or Euclidean) modes, with the two regions being separated by light cone scattering from the BH potential~\cite{Arnaudo:2025uos,Arnaudo:2025kit,Arnaudo:2026tcy}. The systematic evaluation of EFs for KdS QNMs carried out in this paper is a necessary first step to study the complete Green's function for rotating (KdS) BHs. Interesting future extensions of the present work could include (i) a calculation of KdS EFs for Matsubara modes, and (ii) a systematic investigation of the Green's function, including the (numerically challenging) limit where $M^2\Lambda \ll 1$.

Such a study would lead to interesting insights into the different causal contributions to the Green's function of both asymptotically de Sitter and asymptotically flat rotating BHs. In the Kerr limit, the results of Ref.~\cite{Motohashi:2026mbn} show that the Matsubara pole contributions must cancel in the ratio of connection coefficients entering the decomposed Green’s function contribution.

\acknowledgments
We thank N. Speeney and M. Della Rocca for useful comments and discussions.
N.O. was supported by Japan Society for the Promotion of Science (JSPS) KAKENHI Grant No.~JP23K13111 and No.~JP26K17142.
E.B. is supported by NSF Grants No.~No.~AST-2513337, No.~AST-2307146, No.~PHY-2513337, No.~PHY-090003, and No.~PHY-20043, by NASA Grant No.~21-ATP21-0010, by John Templeton Foundation Grant No.~62840, by the Simons Foundation [MPS-SIP-00001698, E.B.], by the Simons Foundation International [SFI-MPS-BH-00012593-02], and by Italian Ministry of Foreign Affairs and International Cooperation Grant No.~PGR01167. Part of this work was carried out at the Advanced Research Computing at Hopkins (ARCH) core facility (\url{https://www.arch.jhu.edu/}), which is supported by the NSF Grant No. OAC-1920103. This project has received funding from the European Union’s Horizon Europe research and innovation programme under the Marie Skłodowska-Curie Staff Exchanges grant agreement No. 101299389 (STRONG).

\appendix
\section{Continued fraction method}
\label{appendix:continued_fraction_method}

The KdS generalization of the radial and angular Teukolsky equations can be cast in the form of Heun differential equations~\cite{HeunDiffEq,Suzuki:1998vy} that can be solved to compute QNMs with the continued fraction method~\cite{Leaver:1985ax,Yoshida:2010zzb}.

Consider the Heun differential equation~\eqref{eq:HeunEq} in its general form
\begin{equation}
\label{eq:HeunEq_repeated}
\begin{split}
    \Bigg\{\frac{\partial ^2}{\partial z^2} \, + \, \left(\frac{\gamma }{z}+\frac{\delta }{z-1}+\frac{\epsilon }{z-z_a}\right)\frac{\partial}{\partial z}\, + \, \\
    \frac{\alpha \beta z-q}{z (z-1) (z-z_a)}\Bigg\}y(z) =0\,,
\end{split}
\end{equation}
where the parameters $\{\alpha,\,\beta,\,\gamma,\,\delta,\,\epsilon,\,q,\,z_a\}$ satisfy the condition $\alpha + \beta +1=\gamma + \delta + \epsilon $. Eq.~\eqref{eq:HeunEq_repeated} has four regular singular points at $z=0,\,1,\,z_a,\,\infty$.

It can be shown~\cite{Yoshida:2010zzb} (see also~\cite{HeunDiffEq,Suzuki:1998vy}) that the expansion
\begin{equation}
\label{eq:HeunEq_sol_decomposition}
    y(z)=\sum_{n=-\infty}^{+\infty}c_n z^n\,,
\end{equation}
is a solution of Eq.~\eqref{eq:HeunEq} if and only if the coefficients $c_n$ satisfy the three-term recurrence relation
\begin{equation}
    \label{eq:3term_rec_rel_Heun}
    \alpha_n c_{n+1}+\beta_n c_{n} + \gamma_n c_{n-1}=0\,,
\end{equation}
with
\begin{subequations}
\label{eq:recurrence_rel_Heun}
\begin{align}
     \alpha_n&= z_a(n+1) (\gamma +n)\,,\\
     \beta_n & = -n^2 (z_a+1)-\\ \nonumber & \quad n (\gamma +z_a (\gamma +\delta -1)+\epsilon -1)-q\,, \\
     \gamma_n&=(n-1) (\gamma +\delta +n+\epsilon -2)+\alpha\beta\,.
\end{align}
\end{subequations}

To simplify the notation, in the following we will drop the $(\ell\,,m)$ subscripts for the solutions of the Teukolsky equations used in Secs.~\ref{sec:intro} and~\ref{sec:theory}.

\subsection{Solution of the recurrence relations for the Teukolsky equations}

Once the solutions of the radial and angular Teukolsky equations are cast as recurrence relations of the form~\eqref{eq:3term_rec_rel_Heun}, they can be solved simultaneously by imposing the continued fraction conditions~\cite{Leaver:1985ax}
\begin{align}
    0&=\beta_0^r-\frac{\alpha_0^r\gamma_1^r}{\beta^r_1-}\frac{\alpha_1^r\gamma_2^r}{\beta^r_2-}...\,,
\label{eq:continued_fraction_radial}\\
    0&=\beta_0^\theta-\frac{\alpha_0^\theta\gamma_1^\theta}{\beta^\theta_1-}\frac{\alpha_1^\theta\gamma_2^\theta}{\beta^\theta_2-}...\,.
\label{eq:continued_fraction_angular}
\end{align}
The iterative simultaneous solution of these equations provides the QNM frequencies $\omega$ and the angular eigenvalues ${\lambda}$.

\subsubsection{Angular equation}

The angular Teukolsky equation~\eqref{eq:angular_Teukolsky} can be cast in the form~\eqref{eq:HeunEq_repeated} with the transformation~\cite{Suzuki:1998vy}
\begin{equation}
    \label{def: angularFuncTransformation}
    S(z)=z^{A_1}(z-1)^{A_2}(z-z_a)^{A_3}(z-z_\infty)y(z)\,,
\end{equation}
where
\begin{equation}
\label{eq:angular_variable}
    z=\frac{1-\frac{i}{\sqrt{\alpha}}}{2}\frac{x+1}{x-\frac{i}{\sqrt{\alpha}}}\,,
\end{equation}
\begin{align}
    z_a&=-\frac{i(1+i\sqAlpha)^2}{4\sqAlpha}\,,\nonumber\\
    z_\infty&=-{i(1+i\sqAlpha)}/{(2\sqAlpha)}\,,
\end{align}
and
\begin{align}
  A_1&=\pm\frac{m-s}{2}\,, \nonumber \\
  A_2&=\pm\frac{m+s}{2}\,,\nonumber \\
  A_3&= \pm\frac{i}{2}\left(\frac{1+\alpha}{\sqAlpha}a\omega -\sqAlpha m -is\right)\,.
    \label{eq:angular_exponents}
\end{align}
Note that Eq.~\eqref{eq:angular_Teukolsky} has five regular singular points at $x=\pm 1,\,\pm i/\sqrt{\alpha},\,\infty$. One of these, the point 
$x=\infty$, is factored out with the transformation~\eqref{def: angularFuncTransformation}.
The exponents $A_1, A_2, A_3$ in Eq.~\eqref{eq:angular_exponents}, as well as the exponent 1 for the $z-z_\infty$ term, are obtained through the indicial equations that we get when we expand around the corresponding singular point and keep only the most divergent terms.
We choose the positive sign for $A_1=\frac{m-s}{2}$ and the negative sign for $A_2=-\frac{m+s}{2}$, as in Ref.~\cite{Hatsuda:2020sbn}.
The continued fraction equation~\eqref{eq:continued_fraction_angular} selects the minimal solutions in the large-$n$ regime, so that the regularity of the solution around $x=1$ ($z=1$) is automatically guaranteed.

The solution is also regular at $x=-1$ ($z=0$) if the parameters of the Heun equation~\eqref{eq:HeunEq} satisfy
\begin{equation}
\begin{split}
    &\gamma=2A_1 +1\,, \quad \delta=2A_2 +1\,, \quad \epsilon=2A_3 +1\,, \\
    &q=\frac{i {\rm \Lambda} }{4 \sqrt{\alpha }}+A_1+\left(m+\frac{1}{2}\right) (A_3-\hat A_3)+\frac{1}{2}\,,\\
    &\rho_\pm=1+A_1 + A_2 + A_3 \pm \hat A_3\,,\\
    &\alpha=\rho_+ \,,\quad \beta=\rho_-\,,   
\end{split}\label{eq:angular_Heun_par_1}
\end{equation}
($\hat A_3=(A_3(\omega^*))^*$) when $m-s\ge0$, and
\begin{equation}
\begin{split}
    &\gamma=1-2A_1\,, \quad \delta=2A_2 +1\,, \quad \epsilon=2A_3 +1\,, \\
    &u=\frac{i {\rm \Lambda} }{4 \sqrt{\alpha }}+A_1+\left(m+\frac{1}{2}\right) (A_3-\hat A_3)+\frac{1}{2}\,,\\
    &\rho_\pm=1+A_1 + A_2 + A_3 \pm \hat A_3\,,\\
    &\alpha=\rho_+ -2A_1 \,,\quad \beta=\rho_--2A_1\,,\\
    &q=-2A_1(z_a\delta+\epsilon)+u\,,
\end{split}\label{eq:angular_Heun_par_2}
\end{equation}
when $m-s<0$, where we have used the connection formulae of the Heun functions~\cite{HeunDiffEq}.

\subsubsection{Radial equation}

The radial equation~\eqref{eq:radial_Teukolsky} can similarly be written in the Heun form through the transformation~\cite{Suzuki:1998vy}
\begin{equation}
\label{eq:radial_eigenfunction_2}
\begin{split}
    R=&z^{B_1}(z-1)^{B_2}\left(z-z_a\right)^{B_3}\left(z-z_{\infty}\right)^{B_4} y(z)\,,
\end{split}
\end{equation}
where
\begin{equation}
\label{eq:zrad_transformation}
    z=\frac{(\rpp-\rmin)(r-\rp)}{(\rpp-\rp)(r-\rmin)}\,,
\end{equation}
\begin{equation}
\label{eq:radial_parameters_to_Heun}
\begin{split}
  z_a&=
      \frac{(\rpp-r_-)}{(\rpp-\rp)}\frac{(\rmm-\rp)}{(\rmm-r_-)}\,, \nonumber\\
  z_\infty&=\frac{(\rpp-r_-)}{(\rpp-\rp)}\,,
\end{split}
\end{equation}
and the exponents determined through the indicial equations are
\begin{align}
\label{eq:Brad_exponents}
\nonumber
    B_1&=\frac{1}{2} \left\{-s\pm i  \left(\frac{2 (\alpha +1) K(\rp)}{{\rm \Delta}_r'(\rp)}-i s\right)\right\}\,, \\
    B_2&=\frac{1}{2} \left\{-s\pm i  \left(\frac{2 (\alpha +1) K(\rpp)}{{\rm \Delta}_r'(\rpp)}-i s\right)\right\}\,,\\ \nonumber
    B_3&=\frac{1}{2} \left\{-s\pm i  \left(\frac{2 (\alpha +1) K(\rmm)}{{\rm \Delta}_r'(\rmm)}-i s\right)\right\}\,,\\
    \nonumber B_4&=2s+1\,.
\end{align}

The radial equation has five singular points at $r=r_\pm,\,r'_\pm,\,\infty$. One of these ($r=\infty$) is factored out using the transformation~\eqref{eq:radial_eigenfunction} and the definition \eqref{eq:zrad_transformation} of the variable $z$, such that the outer horizon $\rp$ corresponds to $z=0$ and the cosmological horizon to $z=1$~\cite{Hatsuda:2020sbn}.

By Eq.~\eqref{eq:radial_eigenfunction}, the coefficients $B_1$, $B_2$ are associated to the event horizon ($z=0$) and to the cosmological horizon ($z=1$), respectively.
In Eq.~\eqref{eq:Brad_exponents}, the $+$ ($-$) sign in $B_1$ corresponds to outgoing (ingoing) boundary conditions at the event horizon, while the $+$ ($-$) sign in $B_2$ corresponds to outgoing (ingoing) boundary conditions at the cosmological horizon.
For simplicity, we fix both $B_1$ and $B_2$ to have the $+$ sign: cf.~Eq.~\eqref{eq:B1_B2}. The continued fraction equation~\eqref{eq:continued_fraction_radial} will select the minimal solution that is outgoing at the cosmological horizon. The asymptotic ingoing behavior at the event horizon is guaranteed by defining the parameters of the Heun equation~\eqref{eq:HeunEq} as

\begin{align}
    \alpha&=1+s-2B_1\,, \quad \beta=1 -2B_1- 2 B(\rmin)\,, \nonumber \\ 
    \gamma&=1-s-2B_1\,, \quad \delta=2B_2+s +1\,, \quad \epsilon=2B_3+s +1\,, \nonumber \\
    v &= \frac{1}{\alpha  (\rmin-\rmm) (\rmin-\rp) (\rp-\rpp)}\left(2 i (\alpha +1) a^4 (2 s+1) \omega \right. \nonumber \\
    & \left. -2 i (\alpha +1) a^3 m (2 s+1)+a^2 ((\rp-\rmin) (\lambda +2 (\alpha -1) s)\right. \nonumber  \\
    & \left. +2 i (\alpha +1) \rmin \rp (2 s+1) \omega )+\alpha  (s+1) (2 s+1) \right. \nonumber\\
    & \left. \times(\rmin-\rp) (\rmin \rp+\rmm \rpp)\right)\,, \nonumber \\
    q&=(z_a \delta +\epsilon)(-s-2B_1)+v\,,
\end{align}
where we have used the connection formulae of the Heun functions to select the solution of the Heun equation that behaves like $\sim z^{-2B_1-s}$ for $z\to0$ (event horizon), as discussed in Sec.~\ref{sec:theory}. 
Either sign of $B_3$ can be used for the calculation of the QNM frequencies.

\begin{figure*}[t]
    \centering
    \includegraphics[width=\linewidth]{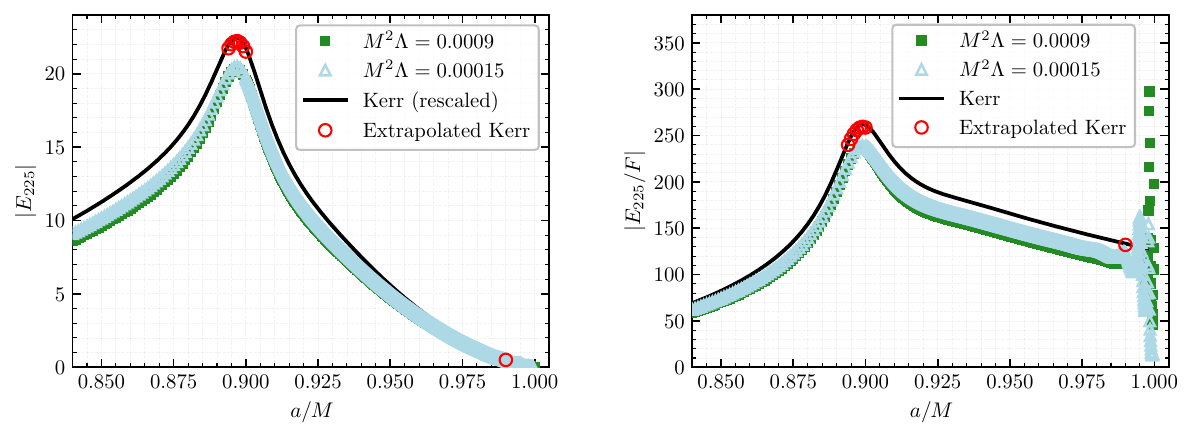}
    \caption{Absolute value of the KdS EFs $|E_{225}|$ as a function of $a/M$ for small but finite $M^2{\rm \Lambda}=0.0009$ (green squares) and $M^2{\rm \Lambda}=0.00015$ (light blue triangles). The black line shows Kerr EFs from Motohashi's catalog~\cite{Motohashi:2024fwt,motohashi_2024_12696857}. The open red dots are Kerr EFs obtained by extrapolating the KdS EFs in the limit $M^2{\rm \Lambda}\to0$ with Eq.~\eqref{eq:fit_model}. In the left panel, the Motohashi Kerr EF values are multiplied by the factor $F$ defined in Eq.~\eqref{eq:EF_conversionFactor}. In the right panel, we divide the KdS EFs by that same factor.}
    \label{fig:convergenceToKerr}
\end{figure*}

\section{Excitation factors with hypergeometric function series}
\label{appendix:STU}
Here we extend the STU/MST method~\cite{Mano:1996vt,Suzuki:1999nn} to compute the KdS EFs~\eqref{eq:excitation_factor} from the solution of the generalized radial Teukolsky equation~\eqref{eq:radial_Teukolsky} as a series of hypergeometric functions. While in the original STU/MST method (and in its extension to KdS~\cite{Suzuki:1999nn}) the solution is expanded in two different sets of functions -- hypergeometric functions near the event horizon, and Coulomb wave functions near the cosmological horizon -- we only expand the solution in terms of hypergeometric functions. This is possible because the solution, with a proper variable redefinition, is regular on the cosmological horizon, and we can always find an overlap region where the solutions at the event horizon and at the cosmological horizon are simultaneously regular.

As shown in Appendix~\ref{appendix:continued_fraction_method}, the radial Teukolsky equation~\eqref{eq:radial_Teukolsky} can be cast into a Heun differential equation~\eqref{eq:HeunEq}, whose solution can be written as a sum of hypergeometric functions~\cite{Suzuki:1998vy,HeunDiffEq}
\begin{align}
\label{eq:hypergeometric_heun_sol}
    y(z)&=\sum_{n=-\infty}^{+\infty} c_n^\nu u_{n+\nu}(z)\,,\\
    u_\nu&=F(-\nu,\nu+w;\gamma;z)\,,
\end{align}
where $w=\gamma+\delta-1=\alpha+\beta-\epsilon$ and $F$ denotes the standard hypergeometric function~\cite{Suzuki:1998vy,HeunDiffEq}. 
This expression depends on the renormalized angular momentum parameter $\nu$, which can be found by solving the three-term recurrence relation~\cite{Suzuki:1998vy,HeunDiffEq}
\begin{equation}
\label{eq:3-term_relation_Hypergeom}
    \alpha^\nu_n c_{n+1}^\nu+\beta_n^\nu c_n^\nu+\gamma_n^\nu c_{n-1}^\nu=0\,,
\end{equation}
where
\begin{widetext}
\begin{subequations}
    \begin{align}
        \alpha_n^\nu&=-\frac{(\nu +n+1) (\delta +\nu +n) (-\alpha+\nu +n+w+1) (-\beta +\nu +n+w+1)}{(2 \nu +2 n+w+2) (2 \nu +2 n+w+1)}\,,\\
        \beta_n^\nu&=\frac{J_n^\nu }{(2 \nu +2 n+w+1) (2 \nu +2 n+w-1)}-z_a(\nu +n) (\nu +n+w)-q\,,\\
        \gamma_n^\nu&=-\frac{(\alpha+\nu +n-1) (\beta +\nu +n-1) (\gamma +\nu +n-1) (\nu +n+w-1)}{(2 \nu +2 n+w-2) (2 \nu +2 n+w-1)}\,,
    \end{align}
\end{subequations} 
and
\begin{equation}
    J_n^\nu=(2 (\nu +n) (\nu +n+w)+\gamma (w-1)) (\alpha \beta +(\nu +n) (\nu +n+w))+\epsilon (\gamma -\delta ) (\nu +n) (\nu +n+w)\,.
\end{equation}
\end{widetext}
This solution is an expansion around the event horizon ($z=0$), but its domain of convergence includes also the cosmological horizon ($z=1$). We will enforce the boundary conditions at both the event and cosmological horizons. In the Kerr case, the outgoing boundary conditions must be imposed at infinity, outside of the domain of convergence, so it is necessary to perform a different expansion of the solution near infinity.

The expansion~\eqref{eq:hypergeometric_heun_sol} is a solution as long as the coefficients of the recurrence relation~\eqref{eq:3-term_relation_Hypergeom} satisfy appropriate continued fraction relations.
The three-term relation~\eqref{eq:3-term_relation_Hypergeom} can be rewritten in terms of the quantities
\begin{align}
    R_n(\nu)=\frac{c_n^\nu}{c_{n-1}^\nu}\,, \qquad L_n(\nu)=\frac{c_n^\nu}{c_{n+1}^\nu}\,,
\end{align}
such that
\begin{align}
\label{eq:RnLn}
R_n(\nu) &= -\frac{\gamma_n^\nu}{\beta_n^\nu + \alpha_n^\nu R_{n+1}(\nu)}\,,\\
L_n(\nu) &= -\frac{\alpha_n^\nu}{\beta_n^\nu + \gamma_n^\nu L_{n-1}(\nu)}\,.
\end{align}
Then equation~\eqref{eq:3-term_relation_Hypergeom} is equivalent to~\cite{Sasaki:2003xr}
\begin{equation}
    \beta_n^\nu+\alpha_n^\nu R_{n+1}(\nu)+\gamma_n^\nu L_{n-1}(\nu)=0\,.
    \label{eq:abc2}
\end{equation}
By solving this equation, we can compute the renormalized angular momentum $\nu$ and the coefficients $c_n^\nu$, and thus the solution $y(z)$ of the radial Teukolsky equation given by Eq.~\eqref{eq:hypergeometric_heun_sol}, which is related to $R(r)$ by Eq.~\eqref{eq:radial_eigenfunction}. 

The coefficients $\alpha^\nu_n$, $\beta^\nu_n$, $\gamma^\nu_n$ of Eq.~\eqref{eq:abc2} depend on the parameters~\eqref{eq:radial_parameters_to_Heun}, and hence on the exponents $B_i$. As discussed in Appendix~\ref{appendix:continued_fraction_method}, with appropriate choices of such exponents we can enforce boundary conditions at either the event horizon or the cosmological horizon.

An expansion similar to Eq.~\eqref{eq:hypergeometric_heun_sol}, performed around the cosmological horizon $z=1$, gives the ingoing and outgoing solutions at $z=1$. 
In this way we can find the solution of Eq.~\eqref{eq:radial_Teukolsky} satisfying ingoing boundary conditions at the event horizon, $y(z)=y^{\text{(in)}}_{\ell m}(z)$.
Near the cosmological horizon, this solution is a combination of solutions with ingoing and outgoing boundary conditions:
\begin{equation}
 y(z)=c^{\text{(out)}}_{\ell m} y^{\rm (up)}_{\ell m}(z) + c^{\text{(in)}}_{\ell m}  y^{\text{(down)}}_{\ell m}(z)\,.
\end{equation}
The coefficients $c^{(\rm out)}_{\ell m}$ and $c^{(\rm in)}_{\ell m}$ of Eq.~\eqref{eq:RinBH} can be computed using the Wronskian method, and the EFs can then be computed as described in Sec.~\ref{subsec:efKdS}.

\section{Excitation factors in the Kerr limit}
\label{appendix:EF_Kerr_limit}
Following Ref.~\cite{Oshita:2021iyn}, the extrapolation is performed with a polynomial model
\begin{equation}
\label{eq:fit_model}
    E^{\text{fit}}_{\ell m}({\rm \Lambda})=\sum_{k=0}^{k_{\rm max}}c_k \left(\sqrt{\frac{\rm \Lambda}{3}}\right)^k\,,
\end{equation}
where $1<k_{\rm max}<N-1$, and $N$ denotes the number of EF data points -- computed for different values of $M^2{\rm \Lambda}$ in the range $0.000125 \leq M^2{\rm \Lambda} \leq 0.0025$ -- used for the extrapolation.

In our case we set $N=100$. We find that the value of $c_0$, corresponding to the Kerr EF, reaches a plateau such that the extrapolated value is stable for $k_{\rm max}\gtrsim 10$. The extrapolation convergences faster when we truncate higher-order terms in the $M^2{\rm \Lambda}\ll1$ expansion of the exponent $B_2$ defined in Eq.~\eqref{eq:defB2}, i.e.,
\begin{equation}
B_2=-i\frac{\rpp^{\,2}\omega}{(\rpp-\rmm)}\,.\label{eq:B2simp}
\end{equation}

In Fig.~\ref{fig:convergenceToKerr} we show the $(\ell,\,m,\,n)=(2,\,2,\,5)$ EFs using two different conventions. In the left panel we rescale the Kerr EFs to match the convention used in this paper, i.e., we multiply them by the factor $F$ defined in Eq.~\eqref{eq:EF_conversionFactor}. In the right panel we divide the KdS values by the same factor $F$ to match the convention used in Ref.~\cite{Kubota:2025hjk}.  The rescaling in the right panel, although equivalent to that in the left panel, shows that our calculation is affected by numerical instabilities in the extremal regime.

\bibliography{biblio}

\end{document}